\documentclass[graybox]{svmult}

\usepackage{type1cm}        
\usepackage{makeidx}         
\usepackage{graphicx}        
\usepackage{multicol}        
\usepackage[bottom]{footmisc}

\usepackage{newtxtext}       %
\usepackage{newtxmath}       

\usepackage{glossaries}      
\usepackage{bm}      

\newacronym{RIXS}{RIXS}{resonant inelastic x-ray scattering}
\newacronym{IXS}{IXS}{inelastic x-ray scattering}
\newacronym{FEL}{FEL}{free-electron laser}
\newacronym{XFEL}{XFEL}{x-ray free-electron laser}
\newacronym{LCLS}{LCLS}{Linac Coherent Light Source}
\newacronym{SASE}{SASE}{self-amplified spontaneous emission}
\newacronym{ARPES}{ARPES}{angle-resolved photoemission spectroscopy}
\graphicspath{{../}{figs/}}

\makeindex             

\begin{document}

\title*{Ultrafast x-ray probes of dynamics in solids}
\author{Mariano Trigo, Mark P. M. Dean, David A. Reis}
\institute{Mariano Trigo \at Stanford PULSE Institute and Stanford Institute for Materials and Energy Sciences (SIMES), SLAC National Accelerator Laboratory, Menlo Park, CA 94025, USA \email{mtrigo@slac.stanford.edu}%
\and Mark P. M. Dean \at Condensed Matter Physics and Materials Science Department, Brookhaven National Laboratory, Upton, New York 11973, USA \email{mdean@bnl.gov}
\and David A. Reis \at Stanford PULSE Institute and Stanford Institute for Materials and Energy Sciences (SIMES), SLAC National Accelerator Laboratory, Menlo Park, CA 94025, USA \email{dreis@stanford.edu}}
%
%
\newcommand\chapternumber{1}
\renewcommand{\theequation}{\chapternumber.\arabic{equation}}
\renewcommand{\thefigure}{\chapternumber.\arabic{figure}}

\maketitle

Advances in our ability to understand and utilize the world around us have always relied on the development of advanced tools for probing and manipulating materials properties.  X-ray matter interactions  played a critical role in the development of the modern theory of solid-state materials that has continued over more than a century.  The development of ever-brighter x-ray sources has facilitated ever more sensitive x-ray scattering and spectroscopy measurements that are able to probe not just the lattice structure, but also the spectrum of elementary excitations in complex materials. The interactions underlying the electronic, magnetic and thermal properties of solids tend to be associated with lengthscales comparable to the atomic separation and  timescales ranging from femtosecond to picoseconds.  The short wavelength, femtosecond pulses from \gls*{XFEL}s  therefore offer unprecedented opportunities to probe and understand material properties on their natural lengthscales and timescales of these processes. This chapter reviews a number of exemplary \gls*{XFEL}-based  experiments on lattice and electronic dynamics, and their microscopic interactions  both near and out of equilibrium.  We conclude with a brief discussion of new forms of spectroscopy enabled by the combination of high flux and short pulse duration and give an outlook for how the field will develop in the future.

\section{Introduction} 
\label{sec:intro}

X-ray radiation  has proven  invaluable for studying the structure and dynamics of matter on the atomic-scale.  A variety of ever-more-sophisticated techniques have been developed over the past hundred and twenty-five years concomitant with improvements in x-ray sources, optics and detectors.  These methods make use of the short wavelength (comparable to the size of atoms), atomic and chemical specificity, as well as the penetrating power of x rays.  Each new advance has led to important discoveries across a wide range of disciplines.  In the case of condensed matter physics, the high brightness and tunability of storage-ring based synchrotron radiation sources enabled the development of ultra-sensitive x-ray methods including non-resonant \gls*{IXS}~\cite{krisch2007inelastic, Baron2015} and \gls*{RIXS}~\cite{Ament2011resonant, Dean2015insights}.  These techniques have been transformational for the study of collective modes of quantum materials.  Here dedicated instruments have been developed that push the resolution in energy and momentum that allow for the study of small crystals and heterostructures, in a variety of environments.  The use of such sources for studying the nonequilibrium dynamics of quantum materials in the time-domain is limited by the pulse duration of the sources which are typically tens to a hundred picoseconds per burst.
However, dynamics at the inter-atomic-scale, including the highest frequency vibrations and the making and breaking of chemical bonds occurs on a few femtosecond time-scale, while the dynamics associated with electron correlation and exchange interaction can be even faster. 

Extremely bright femtosecond \gls*{XFEL} sources  are transforming our ability to follow the dynamics of solid-state materials on the relevant timescales of their low-lying elementary excitations, as well as to study their non-equilibrium behavior on the atomic-scale\cite{wall2016recent,Lindenberg2017,Buzzi2018probing,Dunne2018x-ray,Cao2019ultrafast}.  The first hard x-ray \gls*{XFEL}, the \gls*{LCLS} became operational a decade ago, initially lasing at a wavelength of $1.5$~\AA, immediately exceeding the peak brightness of the most intense x-ray light sources in operation by 9--10 orders of magnitude~\cite{Emma2010first}. This dramatic increase in peak brightness was due to a combination of the ultrashort pulse duration of $\sim 100$~fs,  large pulse energy $\sim$~mJ and near-transform-limited transverse spatial coherence characteristic of the high-gain \gls*{SASE} process~\cite{Kondratenko1980generating, Bonifacio1984collective,Bonifacio:1984aa}.   Since this time a number of hard x-ray \gls*{FEL} facilities have been built 
around the world~\cite{Ishikawa2012compact, Abeghyan2019, Kang2017hard, milne2017swissfel}.  The capabilities and performance of 
the various \gls*{XFEL}s  has also grown including self-seeding~\cite{Amann2012demonstration,Inoue2019}, shorter pulses of a few tens of femtoseconds down to sub-femtosecond (attosecond) regime~\cite{Huang2017,Duris2020}, tunability of x-ray energy up to $\sim$20~keV~\cite{Tono_2013}, 
as well as two-color operation~\cite{Hara:2013aa}.  The next generation superconducting LINAC based \gls*{XFEL}s will operate at much higher average power and repetitions rates \cite{galayda2018, zhu2017sclf}.

We will not discus the physics of the FEL process here, as there are a number of excellent references for the interested reader~\cite{Huang2007review,saldin2010statistical,McNeil2010,Pellegrini2016pellegrini,seddon2017}. Instead we will focus on describing a sub-set of experiments that make use of the unique properties of the \gls*{XFEL} for studying the ultrafast dynamics of quantum materials.  After a brief discussion of the general experimental considerations, we focus on discussing different classes of dynamics in optical laser-pumped materials probed by both non-resonant and resonant scattering techniques.
We begin by considering examples where non-resonant time-resolved diffraction probes coherent excitation of lattice modes and electron-phonon coupling in the linear-response regime, both at zone center and in the case of diffuse x-ray scattering, for two-phonon excitations spanning the Brillouin zone.  We then go on to discuss examples where these methods have been used to uncover the nonlinear couplings between multiple degrees of freedom, as well as cases where photoexcitation induces excitations well-beyond linear response  using non-resonant diffuse scattering to look at metal-insulator transitions and resonant diffraction to study changes in charge and orbital ordering. Finally we describe experiments where time-resolved resonant inelastic scattering provides combined momentum and energy-resolved information on magnetic, orbital and charge excitations. 
We have not attempted to  be comprehensive and notably have left out a number of important applications and techniques including lensless imaging~\cite{Nugent2010coherent}, x-ray photon correlation spectroscopy \cite{Shpyrko2014xray} and (for the most part) x-ray nonlinear optics in solids \cite{Glover2012,Shwartz2014,Tamasaku:2014qf,Fuchs2015,Ghimire2016,Fuchs2018Young}. 

Finally we note that even before the \gls*{XFEL}, femtosecond duration x-ray pulses were produced in laser-plasma interactions~\cite{Murnane1991ultrafast,Sokolowski_Tinten_2004} laser-slicing on storage rings~\cite{Schoenlein2000generation} and via electron compression in linear-accelerator based spontaneous synchrotron sources~\cite{bentson2003, Lindenberg2005,Cavalieri2005}.  The relatively low flux and availability of these sources have limited their utility. Nonetheless these sources were  critical in developing the nascent field of ultrafast x-ray science ~\cite{Rousse2001rmp,Cavalleri2002Review,Reis2006ultrafast,bargheer2006recent} including the experiments described here. 

\section{X-ray free-electron laser-based pump-probe methods \label{sec:sources}}

The different experiments discussed in this chapter on solid-state dynamics all make use of the combination of the short pulse duration and high flux available on the \gls*{XFEL}.  
 In particular, each derives its temporal-resolution through stroboscopic, pump-probe, methods, whereby a femtosecond long-wavelength pump (often a near-IR laser) excites the material which is subsequently probed by the x-rays captured on a relatively slow detector.  In this manner the dynamics are built up statistically as a function of the relative timing of the pump and probe, and the resolution is given by the combination of the pump duration, the probe duration, and our ability to control (or measure) the relative delay.  The earlier \gls*{XFEL} experiments were limited by  both the timing jitter and drift between the optical and x-ray laser, although subsequently nearly pulse-length limited resolution has been achieved using a single-shot x-ray arrival time monitor~\cite{Bionta2011spectral, chollet2015}.  
In addition, because the SASE process starts up from noise all the various properties of the x-ray pulse  fluctuate from shot-to-shot, such that in addition to the timing, a typical experiment needs to monitor the pulse properties and read the detectors on a shot-by-shot basis, at a few tens to 120~Hz repetition rate.  Here care needs to be taken to minimize systematic errors over the course of an experiment, as well as to handle the processing and analysis of the data (which can easily exceed several TB/hour).

\section{Coherent phonon spectroscopy in linear response}\label{sec:coherent_phonons}

We begin our discussion by considering time-domain measurements of coherently excited collective lattice modes (phonons). Coherent phonon spectroscopy~\cite{dhar1994timeresolved,merlin1997generating} has been an invaluable tool in the optical domain for probing the vibrational properties of solids. An attractive feature of time-domain, coherent phonon spectroscopy is that, similar to nuclear magnetic resonance (NMR), the time-domain sampling of the oscillations produced by a coherent vibration allows for  frequency resolution comparable or surpassing the natural oscillator linewidth.  X-ray scattering has several advantages over optical methods for coherent phonon spectroscopy, including  direct measurements of lattice displacements and momentum resolution that is inaccessible with long-wavelength probing.  Much of the early work on time-resolved x-ray scattering  focused on scattering from coherent acoustic phonons\cite{Rousse2001rmp,Reis2006ultrafast,bargheer2006recent}.  X-ray FELs have the advantage of much better time resolution and flux than any previous source, so it is no surprise that they are powerful tool for measuring coherent phonons with much better sensitivity and from a wider range of materials than previously possible.

%
\begin{figure}[b]
\sidecaption
\includegraphics[width=11.5cm]{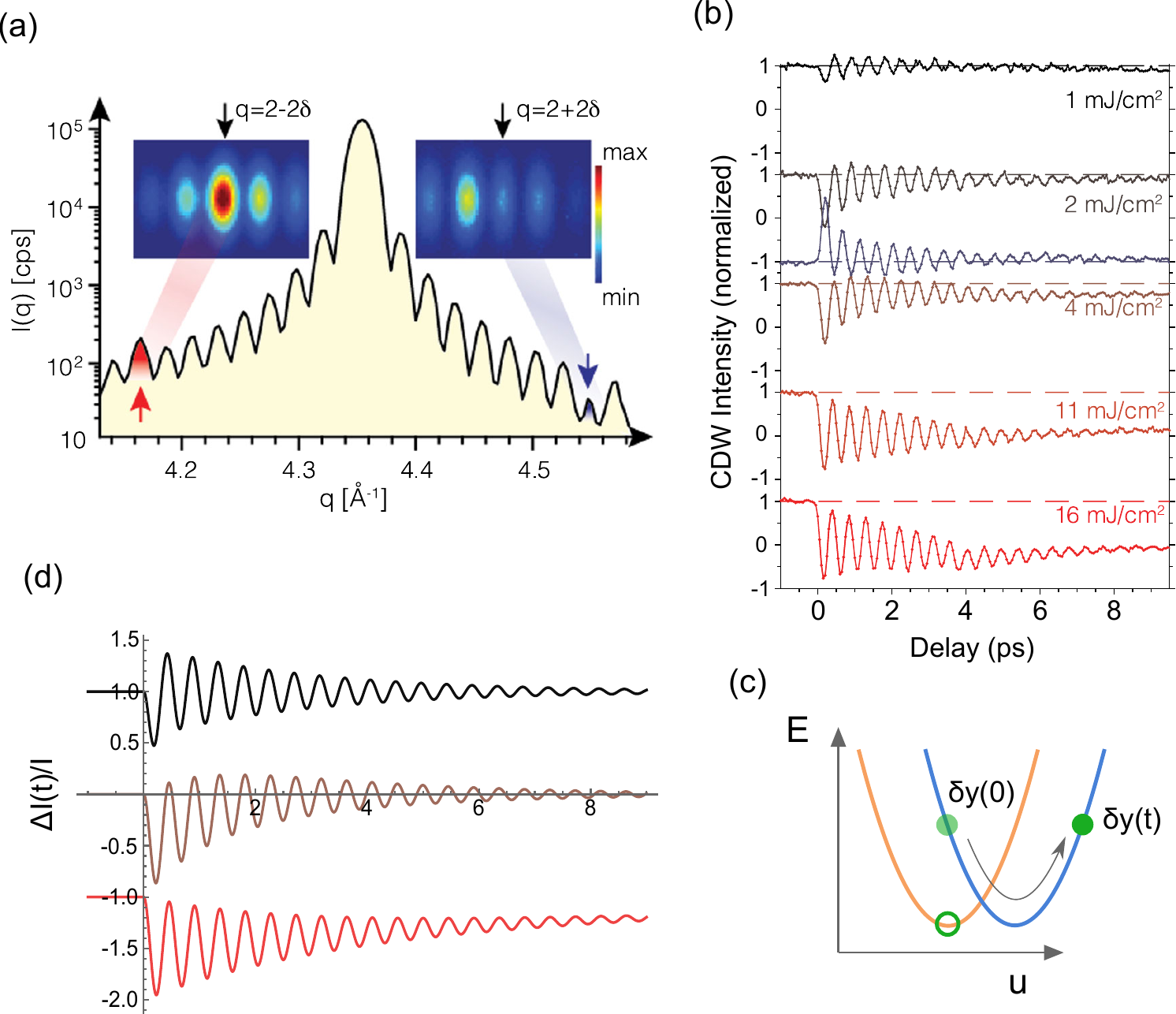}
\caption{Example of coherent phonons probed by ultrafast x-ray diffraction. (a) The diffraction pattern around $Q = (0\ 0\ 2q)$ of a 28~nm-thick Cr film. The small oscillations are sidebands due to finite size. The inset shows a detector image at the CDW wavevector. (b) Dynamics of the normalized CDW intensity for several incident fluences. (c) Schematic of the displacive excitation of a coherent phonon. (d) Calculation of the atomic displacement from Eq.~\ref{eq:decp_full} using parameters from~\cite{Singer2016}. The curves are displaced vertically for clarity. (a-b) Reprinted figure elements with permission from Ref.~\cite{Singer2016} Copyright 1996 by the American Physical Society.}
\label{fig:Singer_CDW}       
\end{figure}
For example, Singer et al.\ used ultrafast x-ray diffraction at the LCLS to probe the dynamics of the charge density wave (CDW) in a 28~nm-thick film of  elemental chromium (Cr)~\cite{Singer2016}. This material exhibits an incommensurate spin density wave (SDW) below $T_N = 290$~K accompanied by a CDW at $2 q$, where $q$ is the magnitude of the SDW wavevector. In the sample studied in~\cite{Singer2016}, the CDW wavevector, $(0\ 0\ 2q)$ was perpendicular to the film. The finite thickness of the film gives rise to finite-size side-bands of the crystal Bragg peak as shown in Fig. \ref{fig:Singer_CDW}(a). These finite-size satellites align and are commensurate with the CDW wavevector due to it being pinned to the film surface.  Importantly, these finite size sidebands at $- 2q$ and $+2q$ interfere constructively (destructively) with the x-ray field scattered by the CDW, resulting in an increase (decrease) of the CDW peak intensity, respectively. This interference, akin to a heterodyne measurement, makes the intensity depend linearly on the atomic displacement associated with the CDW, $\delta y$\cite{kozina2017heterodyne} and the measurement is sensitive to the sign of the atomic motion with $\Delta I(t) = \pm 1$ corresponding to $\delta y = \pm 1$ in normalized units, respectively.
The dynamics of the normalized intensity change for the $\mathbf{Q} = (0\ 0\ k)$ peak, with $k = 2 \pm 2q$  are shown in Fig.~\ref{fig:Singer_CDW}(b) for several incident pump fluences.
The observed oscillations, with period $\sim 0.45$ ps (frequency $2.2$~THz) originate from the amplitude mode of the CDW.

The CDW dynamics in Fig.~\ref{fig:Singer_CDW}(b) follow the displacive excitation of coherent phonons (DECP)~\cite{Zeiger1992theory,merlin1997generating} common among many charge-ordered systems under ultrafast excitation.  In DECP, the coherent motion of the mode is initiated by a sudden displacement of the parabolic potential as shown in Fig.~\ref{fig:Singer_CDW}(c). The general solution for the atomic displacement $\delta y(t>0)$ under a displaced harmonic potential  is\cite{Zeiger1992theory}
\begin{equation}\label{eq:decp_full}
    \delta y(t) = \frac{A}{2}\frac{\Omega^2}{\Omega^2 + \beta^2 - 2 \gamma \beta}\left\{ e^{-\gamma t}\left( \cos \Omega t - \frac{\beta - \gamma}{\Omega} \sin  \Omega t\right) - e^{- \beta t} \right\},
\end{equation}
where $\Omega$, $A$, and $\gamma$ are the frequency, amplitude and damping constant of the oscillator, and $\beta^{-1}$ is the time scale for return to equilibrium of the displaced harmonic potential.
If $\beta - \gamma \ll \Omega$, i.e. a harmonic system with a slow recovery of the potential compared to the frequency, Eq. \ref{eq:decp_full} can be simplified to
\begin{equation}\label{eq:decp_simple}
    \delta y(t) = \frac{A}{2} e^{-\gamma t}\left( \cos \Omega t - e^{-(\beta - \gamma) t}  \right)
\end{equation}
which is an oscillatory (cosine) motion with a slowly recovering equilibrium offset. In the limit of slow relaxation of the potential $\beta \approx 0$ and  $\delta y(t) = \frac{A}{2} (e^{-\gamma t} \cos \Omega t - 1) $ corresponding to a cosine-like phase.

The opposite limit where the displaced potential recovers quickly is $\beta \gg \gamma$ and we get
\begin{equation}\label{eq:decp_impulsive_limit}
   \delta y(t) = \frac{A}{2} \frac{\Omega \beta}{\beta^2 + \Omega^2 - 2 \gamma \beta} e^{-\gamma t} \sin \Omega t.
\end{equation}
In this case the oscillatory motion is sine-like
which behaves similarly to that achieved by an impulsive excitation. This is the behavior of the low-fluence trace from~\cite{Singer2016} shown in Fig.~\ref{fig:Singer_CDW}.
In Fig.~\ref{fig:Singer_CDW}(d) we show the normalized dynamics (i.e. setting the amplitude $A = 1$ for all traces for clarity) of $\delta y(t)$ predicted by Eq.~\ref{eq:decp_full} with parameters corresponding to fluences of 1, 4 and 11~mJ/cm$^2$~(see Table S1 in the Supplementary Material of \cite{Singer2016}) [same color code as Fig.~\ref{fig:Singer_CDW}(b)]. The crossover between sine-like and cosine-like motion is clear as the fluence increases and $\beta$ decreases.

\begin{figure}[htb]
\sidecaption[t]
\includegraphics[width=7.5cm]{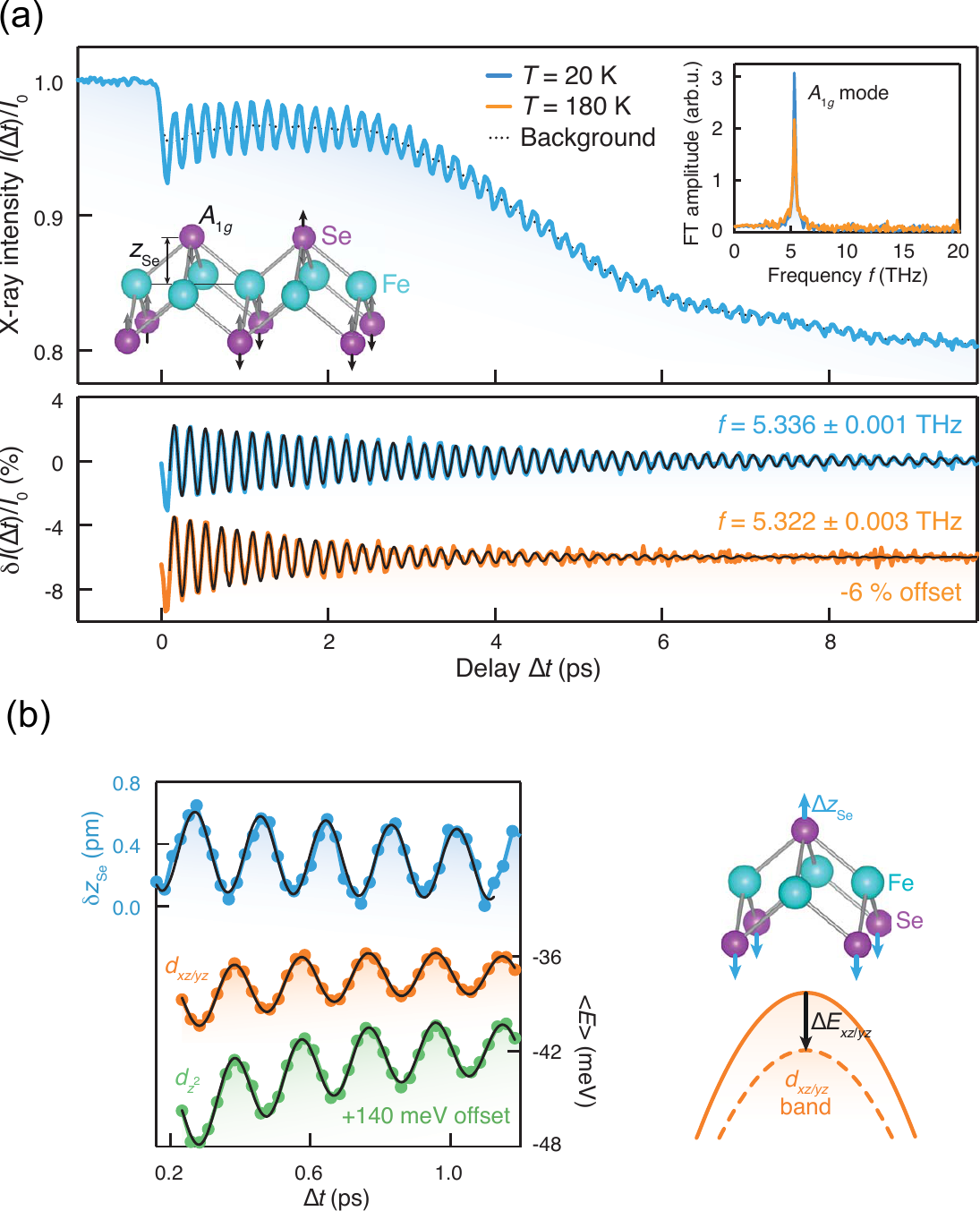}
\caption{ Ultrafast x-ray diffraction from coherent phonons in FeSe. (a) The top panel shows the integrated intensity of the $(0\ 0\ 4)$ Bragg peak after ultrafast excitation with 1.55~eV photons. The lower panel shows the oscillatory component with a slowly varying background subtracted. The inset in the top panel shows the Fourier transform of the oscillatory component in the lower panel. (b) The calibrated $\Delta z_\mathrm{Se}$ (top curve, left y-axis); and corresponding $\Delta E_{xz/yz}$ and $\Delta E_{z^2}$ extracted from time resolved ARPES (middle and lower curves, right y-axis). Also shown is a schematic of the motion and the electronic band displacement. Adapted from~\cite{Gerber2017femtosecond}. Reprinted with permission from AAAS.}
\label{fig:coherent_phonons_gerber}       
\end{figure}

Generally, under the influence of a coherent phonon the electronic bands shift in energy due to their  coupling with the lattice, quantified by the deformation potential. Gerber et al. ~\cite{Gerber2017femtosecond} used time- and \gls*{ARPES} 
to measure $\Delta E_{xz/yz}$ and $\Delta E_{z^2}$, the shift in the $d_{xz/yz}$  and $d_{z^2}$ bands of FeSe due to a coherent lattice vibration. They used femtosecond hard x-ray pulses from the LCLS to measure precisely the displacement of the Se atom $\Delta z_{\mathrm{Se}}$ associated with the coherent $A_{1g}$ phonon mode induced by the pump pulse which modulates the electronic bands.
In their diffraction experiment, the phonon dynamics were excited by an IR laser with 1.55~eV photons and 40~fs in duration and the diffraction of the $(0\ 0\ 4)$ Bragg peak was probed by a delayed 8.7~keV x-ray pulse of 25~fs duration from the LCLS.
Figure \ref{fig:coherent_phonons_gerber}(a) shows the dynamics of the integrated intensity normalized to the laser-off intensity when pumped with a fluence of 1.83~mJ/cm$^2$. The oscillations are due to a 5.3~THz $A_{1g}$ phonon coherently modulating the $z$ position of the selenium atoms,  depicted in the inset of the top panel. The spectrum in the inset shows the Fourier transform of the background-subtracted traces in Fig.~\ref{fig:coherent_phonons_gerber}(a). Using information on the equilibrium atomic coordinates and the $A_{1g}$ mode eigenvector, they relate the measured intensity to the absolute atomic motion involving the Se atom away from the Fe, which at this fluence is $\Delta z_{\mathrm{Se}} \sim 2$~pm near $\Delta t = 0$~ps [see top curve in Fig.~\ref{fig:coherent_phonons_gerber}(b)].
In their time-resolved \gls*{ARPES} measurement, Gerber et al.\ observed that the $d_{z^2}$ and $d_{xz/yz}$ bands shifted downwards in energy as the Se atom moved away from the Fe atom. In Fig.~\ref{fig:coherent_phonons_gerber}(b) we show $\Delta z_\mathrm{Se}$ (top curve, left y-axis) and $\Delta E_{xz/yz}$, $\Delta E_{z^2}$ (middle and lowest curve, right y-axis), as well as the schematic of the corresponding atomic and band  motion.
A careful check of the linear fluence dependence ensures that the measurements are representative of the equilibrium properties. Thus by fitting the linear fluence dependence of both measurements they were able to obtain the deformation potentials of the two bands $\Delta E_{xz/yz}/\Delta z_{Se} = -13 \pm 2.5$ and $\Delta E_{z^2}/\Delta z_{Se} = -16.5 \pm 3.2$~meV/pm. These values are significantly higher than those predicted by DFT calculations~\cite{Gerber2017femtosecond}. Their interpretation 
is that electronic correlations enhance the electron-phonon coupling in FeSe. They argue that such concerted electron-electron and electron-phonon interactions may enhance superconductivity in FeSe, which depends exponentially in the value of the electron-phonon coupling. This work shows that a precise measure of atomic motion in the linear response regime combined with time-resolved \gls*{ARPES} enables model-free measurements of fundamental physical quantities.

The discussion above considers coherent phonons where the macroscopic mode amplitude $\langle u_\mathbf{q}\rangle(t)$, with $u_\mathbf{q}$ the microscopic mode amplitude at wavevector $\mathbf{q}$, oscillates coherently in time at the phonon frequency. Here the angle brackets correspond to ensemble averages, appropriate for macroscopic samples measured with many independent x-ray exposures. These coherent modes can be excited by a long wavelength near-IR pump only at near-zero wavevector $\mathbf{q} \approx 0$ because of the small momentum of the near-IR photons. This is the case in materials with well-defined translational symmetry (continuous or discrete). Alternatively,  in the presence of weak disorder with Fourier component at $\mathbf{q}\neq 0, $ a coherent modes, $\langle u_\mathbf{q}\rangle(t)$ can be generated as crystal momentum only needs to be conserved up to $\hbar \mathbf{q}$. 
This can happen, for example when a static CDW develops, which can provide the momentum to ``fold'' a large wavevector $\mathbf{q}$ back onto the Brillouin zone center, as discussed above.

Optical excitation of a material also produces a different kind of phonon dynamics with the average $\langle u_\mathbf{q}\rangle = 0$ but where the variance $\langle u_{-\mathbf{q}} u_{\mathbf{q}}\rangle(t)$ oscillates in time at ${\emph twice}$ the phonon frequency. Wavevector conservation still applies here, but because this motion involves coherence between $u_{\mathbf{q}}$ and $u_{-\mathbf{q}}$, the total wavevector is still $\mathbf{q} - \mathbf{q} = 0$. In this case, the coherence in the mode variance can span the Brillouin zone. Trigo, Reis and coworkers demonstrated in 2013 that femtosecond x-ray diffuse scattering is sensitive to these dynamics in experiments on near-infrared pumped germanium using the XPP station on LCLS\cite{Trigo2013Fourier}. They demonstrated temporal coherences due to phonons across an extended regions of the Brillouin zone by recording the diffuse x-ray intensity between Bragg peaks. Diffuse intensity, often associated with an unwanted background, contains information on the non-periodicity of a perfect crystal, as the the measured intensity is related to the projection of the mean-square phonon displacements on the momentum transfer; in the case of  non-thermal vibrations it originates from temporal correlations in $\langle u_{-\mathbf{q}} u_{\mathbf{q}}\rangle(t)$ induces by the pump. On the other hand, when the correlations are thermal there is no well defined phase relationship between the modes, and the diffuse x-ray intensity is time-independent \cite{Warren1969Xray}. Instead, in a pump-probe experiment the correlations are time-dependent because the sudden excitation of the crystal induces a well defined phase between  $u_{\mathbf{q}}$ and $u_{-\mathbf{q}}$. 

\begin{figure}[htb]
\includegraphics[width=11.5cm]{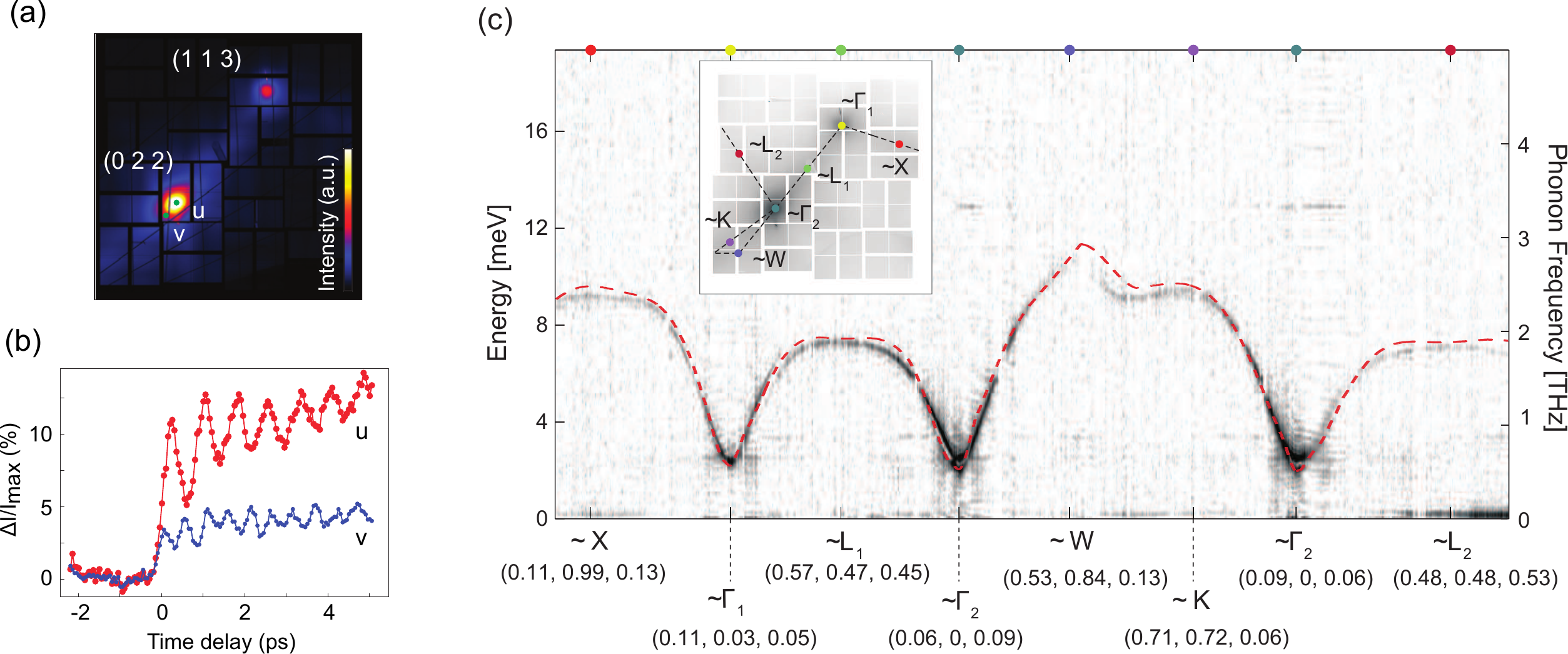}
\caption{ (a) Static thermal diffuse intensity of germanium at room temperature. The two closest reciprocal lattice points are labeled by their Miller indices in the conventional cubic basis. (b) Oscillatory dynamics of germanium after photoexcitation for representative pixels labeled u and v in (a) [adapted by permission from Springer Nature \cite{Trigo2013Fourier}]. (c) transverse acoustic phonon dispersion along the wavevectors marked on the image on the inset. The closest high-symmetry points are labeled and the corresponding coordinates are shown in reciprocal lattice units. The dashed line shows a calculation of the lowest frequency TA branch in Ge. Reprinted figure with permission from \cite{Zhu2015Phonon} Copyright 2015 by the
American Physical Society.}
\label{fig:trigo_ftixs}
\end{figure}

In Fig.~\ref{fig:trigo_ftixs}(a) we show the x-ray diffuse intensity of a single crystal of germanium with 10~keV x-ray photons at LCLS~\cite{Trigo2013Fourier}. The orientation was such that multiple Brillouin zones were intercepted by the Ewald sphere and covered by the detector. The two brightest spots correspond to regions near the $(0\ 2\ 2)$ and $(1\ 1\ 3)$ reciprocal lattice points (conventional unit cell), with the brightest pixels closer to the corresponding zone-center. Importantly, the geometry was chosen such that the Bragg condition was not satisfied, and the reciprocal lattice points are offset in the direction perpendicular to the image~\cite{Trigo2013Fourier}.
In Fig.~\ref{fig:trigo_ftixs}(b) we show the dynamics of the intensity at the two locations in the image labeled $u$, near, and $v$, away, from the zone center, respectively.
In Fig.~\ref{fig:trigo_ftixs}(c) we show the  dispersion obtained from a Fourier transform of the time-domain dynamics along wavevectors indicated in the inset, with labels indicating the closest high-symmetry point in the Brillouin zone of Ge~\cite{Zhu2015Phonon}. The frequency in Fig.~\ref{fig:trigo_ftixs}(c) is obtained from a Fourier transform of the oscillatory signal and the color intensity is related to the amplitude of oscillation. The number of branches visible in this type of experiment is a function of the strength of the coupling to the pump and the polarization sensitivity of the x-ray scattering to the corresponding motion. The time-resolution also determines the highest frequency that can be sampled. For the same reason, the frequency resolution is governed by the longest delay scanned. In this way sub-meV resolution for the excited phonons was achieved by measuring oscillations up to 10 ps.  The amplitude of the oscillations as a function of $\mathbf{q}$ is related to how strongly the various phonon modes couple to the photoexcited charge density, and thus provide a way to access $\mathbf{q}$-dependent electron-phonon matrix elements.

\section{Couplings among multiple degrees of freedom}

In complex materials, the strong interaction among coupled charge, spin, orbital and lattice degrees of freedom can lead to nearly degenerate phases with different broken symmetries~\cite{Imada1998metal}.  
Optical excitation has been used to alter this delicate balance and produce properties not accessible in equilibrium.
In this section we review how ultrafast x-ray scattering is used to probe this coupling in complex materials.
We focus first on the coupling among phonon degrees of freedom, before moving onto coupled charge and spin collective modes.
While near-IR and optical excitation can transiently modify the electronic distribution of a material, which in turn can change structural and magnetic properties, recent development in mid-infrared lasers enable tuning the central frequency of the incident laser pulse to be in resonance with an optically active phonon in the THz range\cite{Foerst2011nonlinear,Mankowsky2014,Kozina2019}. This provides a potentially  attractive avenue to tune materials properties by interacting directly with the lattice modes.

We consider anharmonic couplings between phonon coordinates $Q_1$ and $Q_2$. We write a generic expansion of the anharmonic potential in powers of  quasi-harmonic eigenmodes as
\begin{equation}
V(Q_1, Q_2) = \sum_{n,m} g_{n,m} Q_1^n, Q_2^m
\end{equation}
where some terms may be forbidden by the crystal symmetry. In the following we discuss experimental manifestations due to anharmonic terms in this expansion achieving mode couplings and structural control.

\begin{figure}[htb]
\sidecaption[t]
\includegraphics[width=11.5cm]{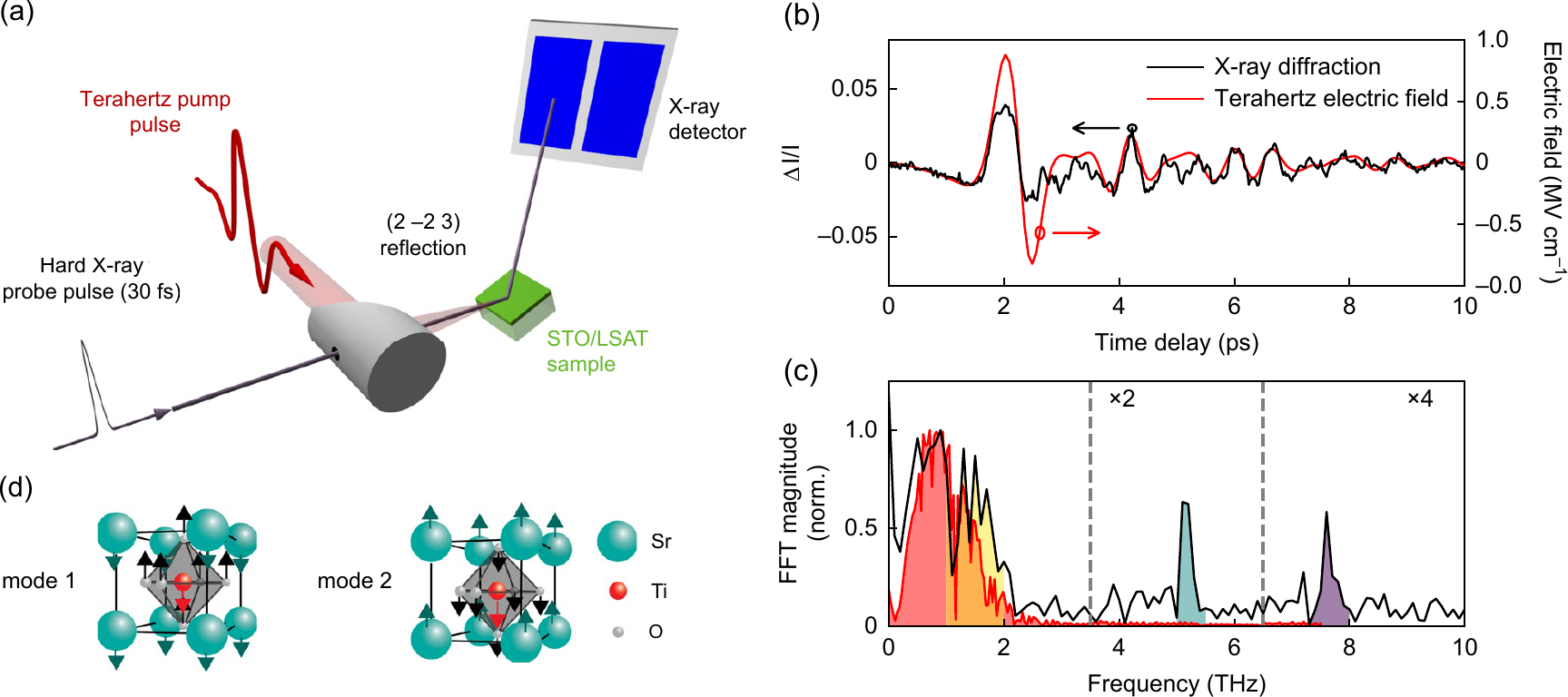}
\caption{ (a) Schematic of the single-cycle THz-pump x-ray-probe experiment on SrTiO$_3$ in Ref.~\cite{Kozina2019}. (b) Dynamics of the intensity of the $(2\ \bar{2}\ 3)$ Bragg peak of SrTiO$_3$  (black curve). Also shown is the electric field of the THz pulse (red curve). (c) Fourier transform of the $(2\ \bar{2}\ 3)$ Bragg peak response. The red and yellow are indicate the spectral regions of the THz field and the soft phonon, respectively. Green and purple indicate the peaks due to anharmonically coupled modes. (d) Illustration of the soft mode (mode 1) and the $\sim 5$~THz mode (mode 2). Adapted by permission from Springer Nature) \cite{Kozina2019}, Copyright (2019).}
\label{fig:kozina_upconversion}
\end{figure}

Kozina et al.\ used such a source of intense THz radiation to drive a particularly anharmonic soft mode in the incipient ferroelectric SrTiO$_3$ into a regime of large amplitude~\cite{Kozina2019}. The large motion induces upconversion into a different phonon mode at higher frequency at the same wavevector\cite{Juraschek2018Sumfrequency}. The displacements internal to the STO unit cell were observed by ultrafast x-ray diffraction at the LCLS\cite{Kozina2019}. 
Fig.~\ref{fig:kozina_upconversion}(a) shows a schematic of their pump-probe experiment. 
A single-cycle THz pulse, with duration $< 1$~ps and spectral content between 0.2 and 2.5~THz is produced by optical rectification of a femtosecond Ti:sapphire pulse in a single crystal of LiNbO$_3$ (not shown), and focused on the sample by an off-axis parabolic mirror. The sample, a 50~nm film of SrTiO$_3$ (STO) on an LSAT substrate is oriented such that the $(2\ \bar{2}\ 3)$ Bragg peak (referred to a cubic Bravais lattice) diffracts the delayed x-ray pulse onto the detector.
Because of its strong anharmonicity the frequency of the STO soft mode is strongly temperature dependent. Kozina et al.\ used this feature to tune the frequency of the mode to the peak of their pump spectrum near $\sim 1$~THz.  Fig.~\ref{fig:kozina_upconversion}(b) shows the dynamics of the integrated intensity of the $(2\  \bar{2}\ 3)$ peak (black curve) together on the measured electric field of the THz field (red curve). The Fourier transform of these traces are shown in Fig.~\ref{fig:kozina_upconversion}(c). The spectrum of both x-ray and THz field has a strong peak at low frequencies $< 1$~THz, indicating that the atomic motion follows the incident field in phase\cite{Kozina2019}. More notably, unlike the THz field, the spectral content of the XRD response has strong features in the range $1 - 2$~THz associated with a strongly damped soft-mode and peaks at $5.15$ and $7.6$~THz corresponding to two additional phonon modes of STO. Since the THz field has no spectral content overlapping these modes, the excitation of these must be due to anharmonic coupling with the soft mode at low frequency.  Also, the $7.6$~THz mode is not IR active, at least in the bulk.
The mechanism was also checked by tuning the mode frequency away from resonance with the THz field by changing temperature. At 250~K the mode frequency is at 2.5~THz and the field does not induce strong anharmonic motion, none of the features are present in the Fourier transform of the x-ray diffraction at this temperature (not shown)\cite{Kozina2019}.

While a full treatment of the potential contains multiple terms, in Ref.~\cite{Kozina2019} they also consider a simplified coupling between the low frequency soft ferroelectric mode $Q_1$ and the $5.15$~THz mode, $Q_2$ that explains the observed dependence on the field strength. Here, $Q_1$ couples directly to the strong THz field as $\sim Z_1 E_{\rm THz}(t)$, with $Z_1$ a mode effective charge, while $Q_2$ couples to $Q_1$ through anharmonic terms in the Hamiltonian $V(Q_1,Q_2) \sim 1/4 g_{4,0} Q_1^4  + g_{3,1} Q_1^3 Q_2$. The equations of motion for this model are
\begin{eqnarray}
\ddot{Q}_1+\gamma_1 \dot{Q}_1+ (\Omega^2_1 + g_{4,0}Q_1^2) Q_1 &=& Z_1 E_{\rm THz}(t) \\
\ddot{Q}_2+\gamma_2 \dot{Q}_2+\Omega^2_2 Q_2 &=& -g_{3,1} Q_1^3,
\end{eqnarray}
where any other nonlinear terms in the equation for $Q_2$ are neglected. Since for stability reasons $g_{4,0} > 0$, the corresponding cubic term in the equation for $Q_1$ makes the frequency higher for large amplitude motion, making it detune from resonance as temperature increases, as observed experimentally. The  term $\sim - g_{3,1} Q_1^3$ in the equation for $Q_2$ produces harmonics of $\Omega_1$, which can overlap with $\Omega_2$ and resonantly drive $Q_2$.

\begin{figure}
\includegraphics[width=\textwidth]{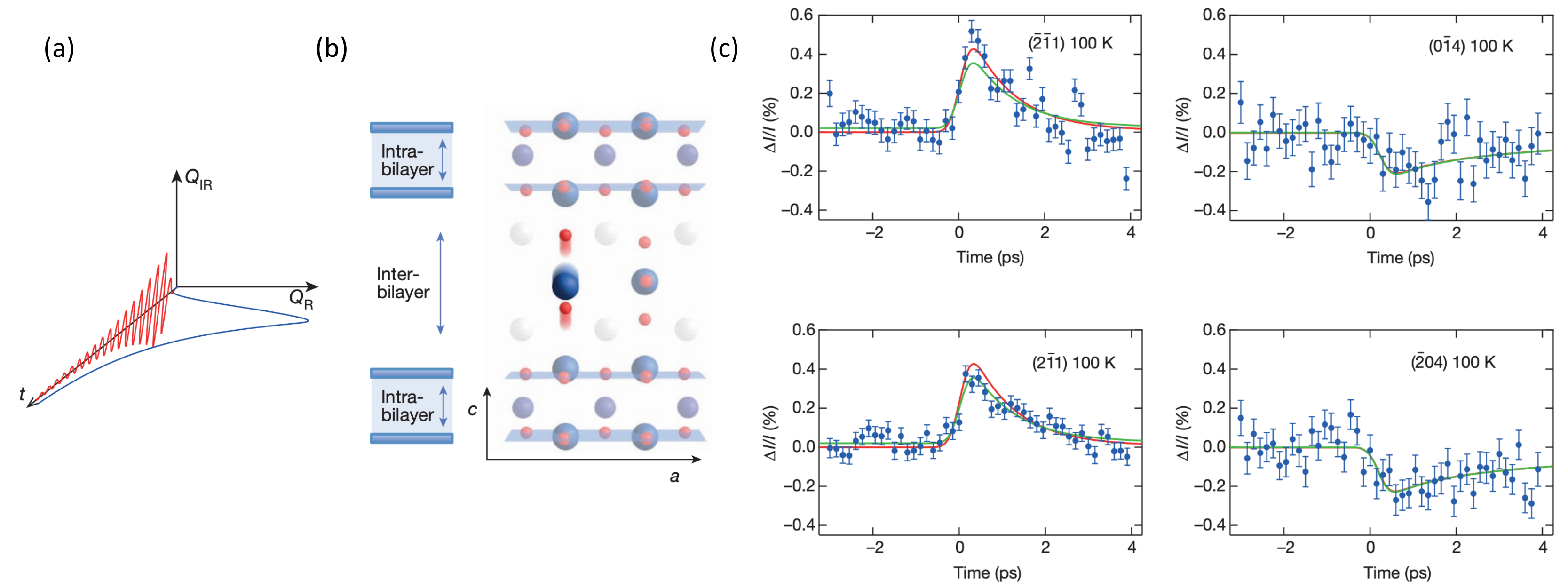}
\caption{Ultrafast modification of the YBa$_2$Cu$_3$O$_{6.5}$ crystal structure after multi-cycle terahertz excitation. (a) Illustration of coupling between infrared and Raman phonon modes through non-linear phononics. (b) Depiction of the intra-bilayer and inter-bilayer distortions in YBa$_2$Cu$_3$O$_{6.5}$. (c) Diffraction intensity of various Bragg peaks. The solid lines represent fits to the data based on the coupling of the $B_{1u}$ phonon to $A_{1g}$ phonons. The green line considers all the $A_{1g}$ phonons; whereas the red line only includes the four most strongly coupled modes. Adapted by permission from Springer Nature \cite{Mankowsky2014}, Copyright (2014).}
\label{fig:YBCO}
\end{figure}

An especially intriguing example of non-linear THz excitation relates to work on the high-temperature cuprates superconductors, whose crystal structures feature stacks of copper-oxide square-net planes. Prior work has studied the transient state of these systems after photo-excitation with radiation centered at $20$~THz ($\sim 15~\mu$m wavelength) detecting increased interlayer transport reminiscent of enhanced superconductivity~\cite{Fausti2011light}. Mankowskii et al.\ have studied YBa$_2$Cu$_3$O$_{6.5}$ after photo-excitation with $c$-axis-polarized THz radiation~\cite{Mankowsky2014}. This radiation was chosen to maximize the interaction with a $B_{1u}$ infrared-active phonon associated with $c$-axis atomic displacements and driven with high fluence in order to maximize non-linear phonon interactions~\cite{Foerst2011nonlinear}.  Figure~\ref{fig:YBCO} illustrate the concepts and data behind this work. The photo-excited $B_{1u}$ mode interacts with $A_{g}$ phonons through non-linear effects and induces changes in the crystal structure, which were probed by ultra-fast diffraction at LCLS through changes in Bragg intensity. By fitting different models that consider these phonon excitations to their data, they conclude that 
the intra-CuO$_2$-plane distance is increased within the transient state at the expense of reducing the inter-CuO$_2$-plane distance. Such a change, would increase the contribution of the Cu $d_{xy}$ orbital to low-energy electronic states, which provides a possible explanation for the proposed enhancement of superconductivity~\cite{Mankowsky2014}.

In general the anharmonic interaction of phonons is not restricted to the coupling between modes at the same wavevector.  A classic example is the anharmonic decay of phonons\cite{Orbach,Klemens} which is the dominant mechanism determining the thermal conductivity of insulators and often plays a central role in structural phase transitions.  In the context of anharmonic decay, first principles density-functional perturbation theory (DFPT) calculations have been used to calculate individual mode couplings for a couple of decades~\cite{Debernardi1995,Debernardi1998,Togo2015Distributions}.  While more traditional inelastic neutron and x-ray scattering methods can give information about the lifetime of modes at a given wavevector, measurements of their underlying microscopic decay channels have not been possible before the advent of the XFELs.

In an experiment by Teitelbaum et al., researchers directly measured the anharmonic decay of a coherent zone center phonon in photo-excited bismuth~\cite{Teitelbaum2018}.  Bismuth is the prototypical Peierls distorted material (which can be viewed as a one-dimensional commensurate charge density wave)  and one of the first materials to show displacive-like excitation of coherent phonons~\cite{Cheng1990,Zeiger1992theory} the dynamics of which at zone-center has been well studied by both ultrafast optical\cite{deCamp2001PRB, hase1996,hase2002,Murray2005,teitelbaum2018real} and ultrafast x-ray diffraction using plasma-sources~\cite{Sokolowski-Tinten2003}, the SPPS linac-based femtosecond spontaneous synchrotron source~\cite{Fritz2007} and laser-slicing in synchrotrons~\cite{Johnson2008nanoscale}. Here the phonon-coordinate for the zone-center, fully-symmetric mode $A_{1g}$ serves as a proxy for the Peierls distortion.  Photoexcitation tends to reduce the bandgap as it lessens the energetic advantage for a structural distortion leading to the displacive excitation of coherent phonons. The decay of the coherent phonon proceeds largely through anharmonic coupling with pairs of acoustic phonons of equal and opposite momenta through a parametric resonance process akin to parametric downconversion~\cite{Fahy2016}.

\begin{figure}[htb]
\sidecaption
\includegraphics[width=7.5cm]{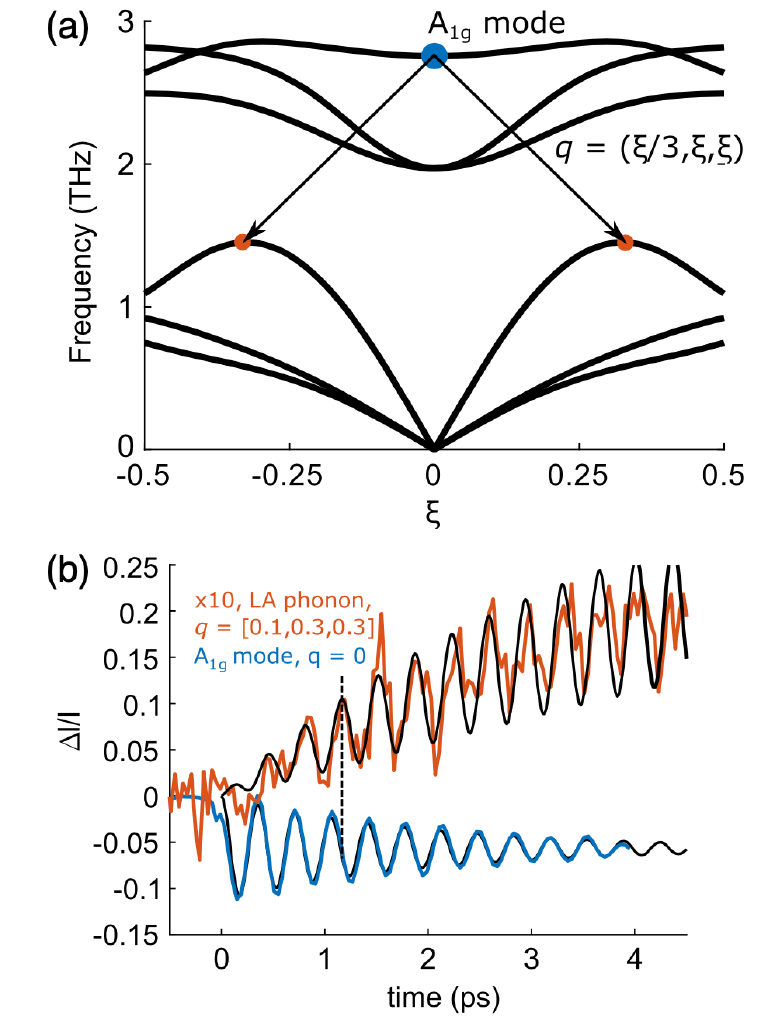}
\caption{Downconversion of $A_{1g}$ mode in bismuth. (a) Phonon dispersion relation in bismuth along the $\bm{q}=(\frac{1}{3}\xi\ \xi\ \xi)$ direction showing one possible decay channel of the $A_{1g}$ mode into a degenerate pair of LA phonons at $\bm{q}=\pm (0.1\ 0.3\ 0.3)$ and (b) the experimental signature corresponding to this particular channel. The blue curve shows the relative intensity change of the $(2\ 3\ 2)$ Bragg peak from which the $A_{1g}$ mode amplitude is measured. The orange
curve shows the relative intensity change of the relevant region of diffuse scattering in the $(0\ 1\ 1)$ zone. The black lines are simulations as described in the paper. Note the dashed
line indicates a $\pi/2$ phase shift between the $A_{1g}$ mode and the target mode which is expected on parametric resonance. Reprinted figure with permission from \cite{Teitelbaum2018} Copyright 2018 by the
American Physical Society.}
\label{f:BiDecay}
\end{figure}

The microscopic decay channels and their coupling constants can be measured directly in the time domain through the build up of squeezed phonon oscillations in femtosecond time-resolved x-ray diffuse scattering measurements when compared to time-resolved diffraction. As described in section \ref{sec:coherent_phonons}, the diffuse x-ray intensity measures the mode variance $\langle u_{-\mathbf{q}} u_{\mathbf{q}}\rangle(t)$, which is driven parametrically by the large amplitude $A_{1g}$ mode. In the x-ray experiments of Teitelbaum et al., the degenerate decay of the $A_{1g}$ oscillations are clearly resolved across a wide-range of the Brillouin zone.  One particular decay channel into two acoustic modes at a specific $\mathbf{q}$ is shown schematically in Fig.~\ref{f:BiDecay}. The coupling constant is determined from the ratio of amplitudes of the squeezed phonon oscillations in $\langle u_{-\mathbf{q}} u_{\mathbf{q}}\rangle(t)$ at a particular wavevector and the amplitude of the $A_{1g}$ coherent phonon oscillation as measured simultaneously in time-resolved diffraction at $\mathbf{q}=0$. In the linear response regime the decay rate of a coherent phonon is expected to be the same as the anharmonic decay rates obtained by perturbation theory, which in the case of bismuth originates from the third-order force constants~\cite{Fahy2016}.  The experimental results give a value for the coupling constant that is within an order of magnitude of DFPT-based calculations, but systematically lower across all the measured channels.  While the reason for the discrepancy is not yet understood the results mark the first momentum-resolved measurement of anharmonic coupling channels and opens the door to similar studies in a broad class of complex materials such as described above, where couplings off-zone center can be critically important.

\section{Beyond linear response}

In this section we describe experiments where the perturbation is pushed beyond linear response. In these experiments, the pump intensity is sufficient to induce non-adiabatic dynamics not accessible thermally, exemplified by lattice symmetry changes. We begin with systems that can be described by a single or few time-dependent order parameter(s). Then we show experiments where fluctuations from this single coordinate along multiple degrees of freedom can be central to the physics of the material, such as the case of VO$_2$.

As one of the simplest examples of dynamics far from harmonic behavior we consider the dynamics of the charge density wave order in SmTe$_3$. In this section we first describe the experiment in SmTe$_3$\cite{Trigo2019Coherent} and introduce the simplest Ginzburg-Landau formalism that results in anharmonic dynamics. We later review experiments where this formalism is extended to multiple degrees of freedom in Pr$_{0.5}$Ca$_{0.5}$MnO$_3$, and finally discuss VO$_2$ where the dynamics cannot be described by a few degrees of freedom but involve a continuum of modes.

\begin{figure}[htb]
\includegraphics[width=11.7cm]{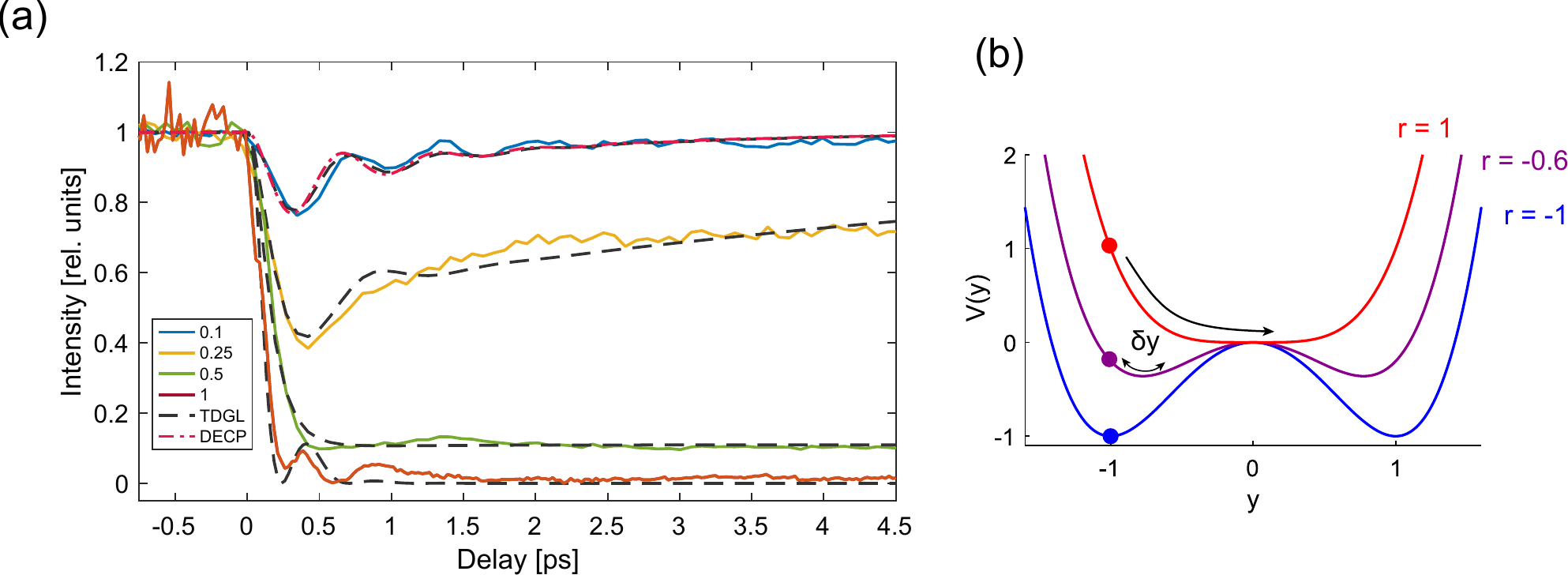}
\caption{ (a) Solid curves: dynamics of the integrated intensity of the $(1\ 7\ q_\mathrm{cdw})$ of SmTe$_3$ at room temperature for several incident pump fluences (in mJ/cm$^2$). Dashed lines are the dynamics computed from the Ginzburg-Landau model described in the text for the corresponding values of $\eta$ (see text). (b) sketch of the effective potential $V(y)$ for several  excitation levels, represented by $r$ in Eq.~\ref{eq:Ginzburg_Landau_potential}. Reprinted figure with permission from \cite{Trigo2019Coherent} Copyright 2019 by the American Physical Society.}
\label{fig:trigo2019}
\end{figure}

SmTe$_3$ exhibits an incommensurate CDW below $T = 416$~K~\cite{ru2008} corresponding to a modulation of the Te-Te planes along the $c$-axis~\cite{ru2008}. The dynamics of the integrated intensity of the $(1\ 7\ q_{\mathrm{cdw}})$  CDW Bragg peak are shown in Fig.~\ref{fig:trigo2019}(a) for several incident fluences\cite{Trigo2019Coherent}. Here $q_\mathrm{cdw} \approx 2/7$ is the CDW wavevector. At low fluence, the response shows oscillations at $1.6$~THz, related to the DECP of the amplitude mode, Eq.~\ref{eq:decp_full}, representing a small deviation from the equilibrium amplitude of the CDW, $\delta y(t)$.
On the other hand, for strong photoexcitation at fluences $\geq 0.5$~mJ/cm$^2$, the peak is strongly suppressed and the dynamics become overdamped and the recover timescale becomes longer. This is a quite common feature in the photoexcited dynamics of broken symmetry states.

At a phenomenological level the dynamics can be described by an extension of the Ginzburg-Landau theory for second order phase transitions~\cite{goldenfeld1992lectures} to the time domain~\cite{Trigo2019Coherent,Schaefer2014Collective,Huber2014Coherent,beaud2014time}. The simplest description considers a scalar order parameter $y$, representing both the amplitude of the CDW lattice distortion and the charge density (or the value of the gap), where in normalized units $|y| \neq 1$ and $y = 0$ represent the distorted low-symmetry phase or undistorted high-symmetry phase, respectively. The potential energy is taken of the form
\begin{equation}\label{eq:Ginzburg_Landau_potential}
    V(y) = V_0 \left( 2 r y^2 + y^4 \right).
\end{equation}
In equilibrium, the parameter $r = T/T_c - 1$, with $T_c$ the transition temperature, controls the relative stability of the two phases. The low-symmetry phase with $y = \pm 1$ is the stable minimum of $V(y)$ for $r = -1$ $(T = 0)$ (Fig.~\ref{fig:trigo2019}(b) (blue curve)). When $r \geq 0$ the potential has a single minimum at $y=0$, and the symmetric phase is stable (Fig.~\ref{fig:trigo2019}(b).
Importantly, to lowest order the intensity of the CDW diffraction peaks is $I_\mathrm{cdw} \propto y^2$. As expected, diffraction alone cannot distinguish between $y = -1$ and $y = +1$, and also $I_\mathrm{cdw} = 0$ for the high-symmetry phase with $y = 0$.

\begin{figure}[htb]
\includegraphics[width=11.7cm]{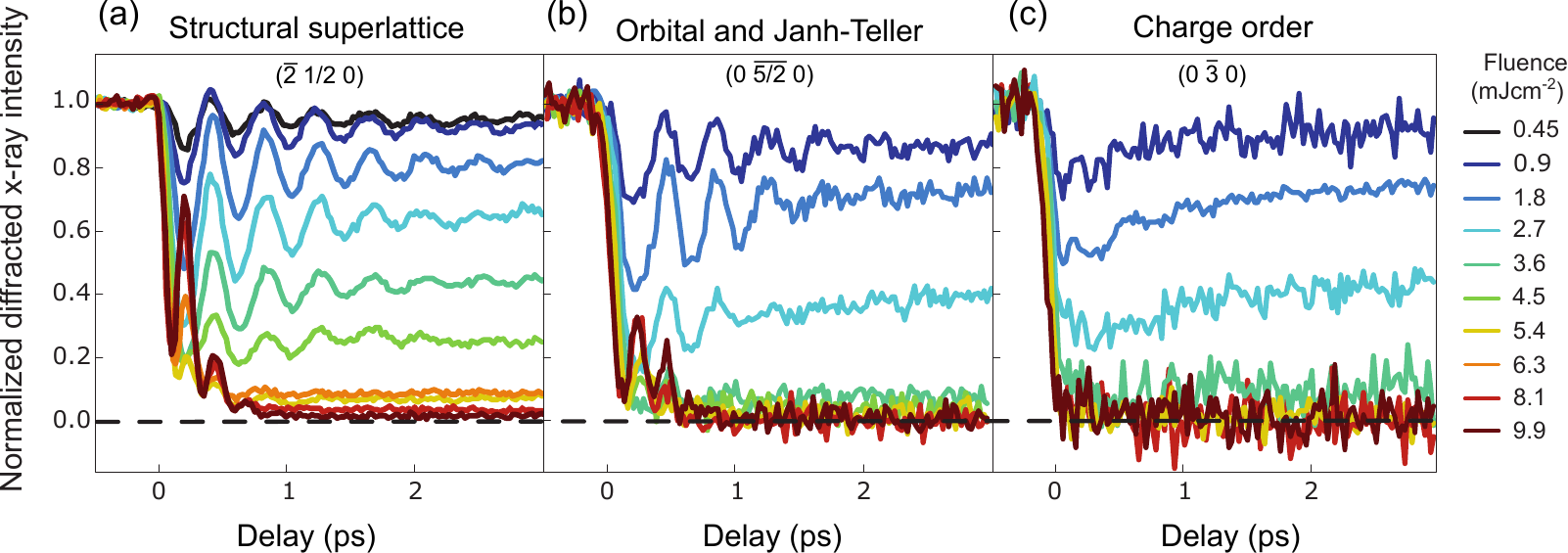}
\caption{ Dynamics of multiple orders in Pr$_{0.5}$Ca$_{0.5}$MnO$_3$ at different incident fluences. (a) Bragg peak from a crystal superstructure. (b) Orbital order peak also associated with a Jahn-Teller distortion of the MnO$_6$ octahedra. (c) charge order peak. Adapted by permission from Springer Nature \cite{beaud2014time}, Copyright (2014).}
\label{fig:beaud2014}
\end{figure}

At ultrafast timescales the temperature is not a well-defined quantity, however a reasonable assumption is that the parameter $r = r(t)$ is related to the photoexcited charge density with a good approximation being $r(t)  = \eta \Theta(t) e^{-t/\tau} - 1$~\cite{Huber2014Coherent}, where  $\Theta(t)$ is a unit step function, $\tau$ is the decay constant of the force and $\eta$ is controlled by the incident pump fluence~\cite{Trigo2019Coherent}.
At room temperature SmTe$_3$ is in the low-symmetry (CDW-distorted) phase and we take $y(t<0) = - 1$ [$y(t<0) = + 1$ is equivalent], representing the static distorted lattice.
After photoexcitation, $r(t>0)  = \eta e^{-t/\tau} - 1$, if the strength of  excitation $\eta$ is small we can consider a small perturbation of the potential $V(y)$, which can be a DECP dynamics for the CDW mode\cite{Trigo2019Coherent}, with a DECP displacement as in Eq.~\ref{eq:decp_full}. At higher fluences (Fig. \ref{fig:trigo2019}(a) green and red traces), the anharmonic behavior is most severe when $\eta \geq 1$, ($r \geq 0$). Here the potential has a single minimum at $y=0$ initially, and the system can reach the symmetric configuration where the distortion associated to $y$ and the diffraction intensity are suppressed completely. This is the essence of the high-fluence behavior in Fig.~\ref{fig:trigo2019} (a) as well as more complex materials like  Pr$_{0.5}$Ca$_{0.5}$MnO$_3$  to be discussed next (shown in Figs.~\ref{fig:beaud2014}(a) and \ref{fig:beaud2014}(b)). More sophisticated models build up from this simple model~\cite{Schaefer2014Collective,Chuang2013RealTime}, but the common feature is some form of time dependent coefficient $r(t)$, representing a phenomenological perturbation to the electronic properties. In all cases, the anharmonic lattice dynamics results from nonlinear (usually quartic) terms in the potential, Eq.~\ref{eq:Ginzburg_Landau_potential}.

The Ginzburg-Landau description can be extended to systems with multiple order parameters, for example representing coupled spin, lattice and charge degrees of freedom~\cite{Chuang2013real}. Beaud and colleagues used the LCLS to probe the dynamics of the phase transformation of Pr$_{0.5}$Ca$_{0.5}$MnO$_3$ whose complex phase diagram is linked to a coupling between lattice, spin, charge and orbital degrees of freedom~\cite{beaud2014time}. Resonant diffraction, where the photon energy is tuned near a core-hole absorption edge (of Mn in this case) can enhance the contribution from valence electrons. This is the basis for novel spectroscopic tools (see more in the resonant inelastic x-ray scattering section \ref{s:RIXS}) making diffraction sensitive to lattice (Fig. \ref{fig:beaud2014}(a)), orbital (Fig. \ref{fig:beaud2014}(b)) and charge order (Fig.~\ref{fig:beaud2014}(c)). Below $5$~mJ/cm$^2$ the charge order is partially suppressed and the orbital and structural superlattice peaks show strong oscillations from a 2.45~THz phonon mode oscillating coherently. At fluence > $5$~mJ/cm$^2$ the three types of order are fully suppressed. Notably, at this fluence the structural and orbital orders exhibit an approximate doubling of the 2.45~THz frequency when the intensity reaches zero, indicative of the $\sim y^2$ scaling of the Bragg intensity when the relevant order parameter approaches the symmetric potential minimum near $y = 0$~\cite{beaud2014time,Huber2014Coherent}.
They model the structural and Jahn-Teller dynamics using a phenomenological Ginzburg-Landau potential similar to the one described above but with four effective lattice degrees of freedom\cite{beaud2014time} representing combinations of phonon modes of Pr$_{0.5}$Ca$_{0.5}$MnO$_3$. Using resonance diffraction to enhance sensitivity to charge order, this experiment showed that although the dynamics are highly out-of-equilibrium, the system has a well defined order parameter across the symmetry breaking phase transition~\cite{beaud2014time}.

In another experiment exploiting resonant scattering to enhance the sensitivity of valence electrons, Lee et al.\ used resonant diffraction to probe the dynamics of charge ordering in La$_{1.75}$Sr$_{0.25}$NiO$_4$ (LSNO)~\cite{WSLee2012}. Unlike the PCMO experiment, where the coherent dynamics can be accounted for by the amplitude of the order parameter alone, in LSNO the pump suppresses the diffraction intensity of the charge-ordered peaks without affecting the peak width. This contrasts with the thermodynamic transition where the peak width (the inverse CDW correlation length) broadens with increasing temperature through $T_c$. This indicates that, for the fluence range studied, the photoexcited transient of the CDW does not contain topological defects~\cite{WSLee2012} in stark contrast to the expected thermal critical behavior. These results in LSNO suggest that photoexcitation populates phase modes at wavevectors other than the nominal ordering wavevector, and these suppress the CDW diffraction peak through a Debye-Waller effect.

\begin{figure}[hbt]
\includegraphics[width=11.5cm]{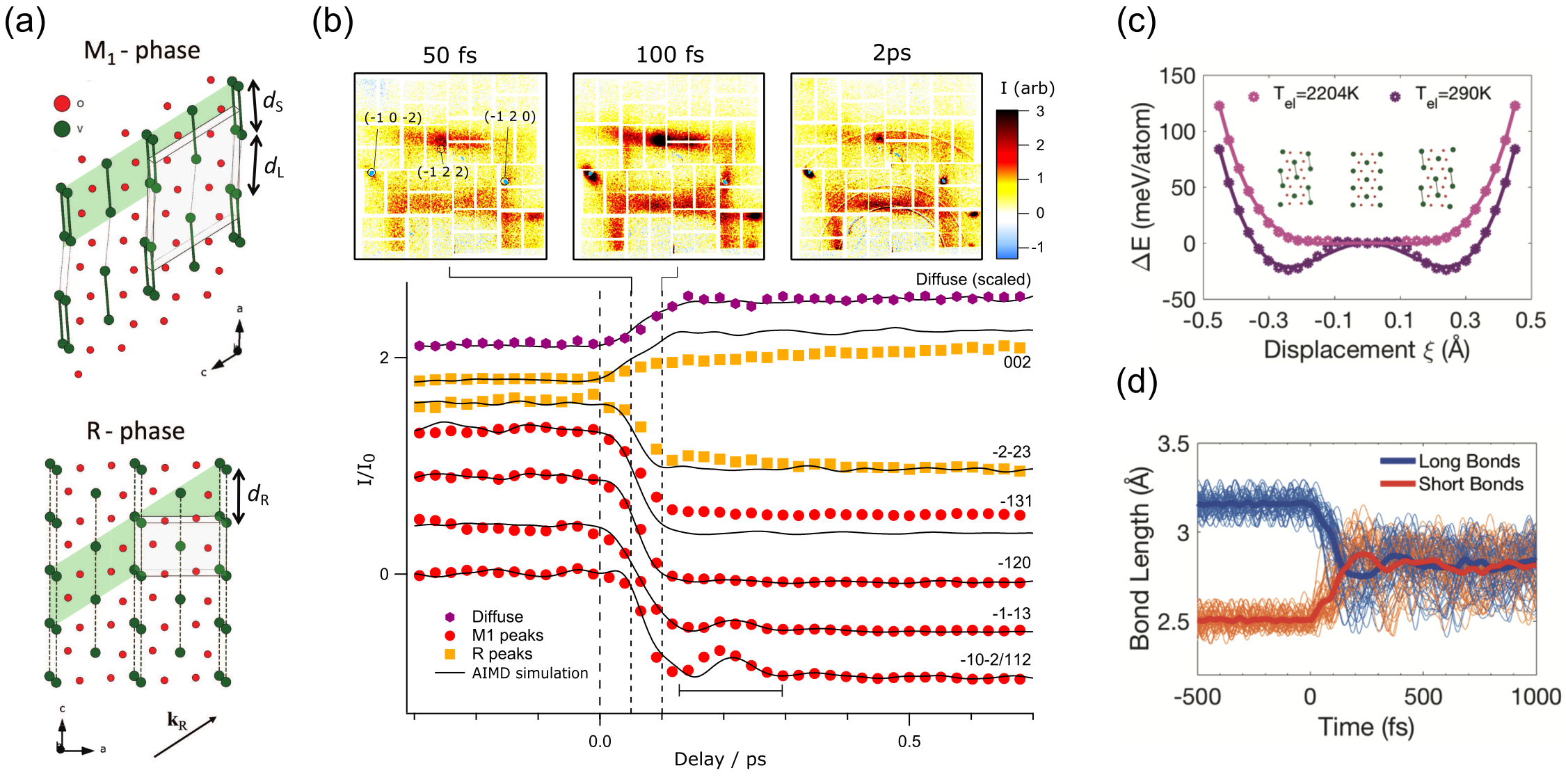}
\caption{ (a) Structure of the $M_1$ and $R$ phases of VO$_2$. The unit cell is indicated with the grey shaded area. (b) top panel: snapshots of the pump-induced change in diffuse intensity $\Delta I(\mathbf{Q},t)$ for representative delays. Bragg peaks corresponding to the $M_1$ cell doubling are indicated ($M_1$ Miller indices). (b) lower panel: dynamics of the integrated intensity of $M_1$ peaks (red symbols), $R$ peaks (orange symbols) and the integrated diffuse intensity (purple symbols). (c) the computed V-V potential for the initial $M_1$ phase ($T_{\mathrm{el}} = 290$~K) and the computed potential under an electronic temperature of $T_{\mathrm{el}} = 2204$~K. (d) AIMD trajectories of the loss of V-V dimerization. From \cite{Wall2018}. Reprinted with permission from AAAS.}
\label{fig:wall2018}
\end{figure}

Just as with any other type of disorder, phase fluctuations should manifest as an increase in the diffuse intensity at Fourier components away from $\mathbf{q}_\mathrm{cdw}$.
Wall et al.\ used diffuse x-ray  scattering at LCLS to probe the photoinduced phase transition of VO$_2$~\cite{Wall2018}. This materials undergoes an insulator-metal transition accompanied by a structural change between monoclinic (M$_1$) and tetragonal structure (Rutile, R)\cite{Morin1959Oxides,Cavalleri2001Femtosecond} at $T_c = 340$~K (Fig.~\ref{fig:wall2018}(a)). The M$_1$ has approximately twice the volume of the Rutile unit cell, and additional commensurate Bragg peaks indicate cell-doubling and long-range ordering of V-V dimers~\cite{Wall2018,Cavalleri2001Femtosecond}. The V-V distances break into long ($d_L$) and short ($d_S$) bonds, which are $d_L = d_S = d_R$ in the R structure, as indicated in Fig.~\ref{fig:wall2018}(a).
The top panel of Fig.~\ref{fig:wall2018}(b) shows representative snapshots of the differential diffuse intensity $\Delta I(\mathbf{Q},t)$ covering multiple Brillouin zones of VO$_2$. Also indicated are sharp, localized features from the $(\bar{1}\ 0\ \bar{2})$, $(\bar{1}\ 2\ 2)$ and $(\bar{1}\ 2\ 0)$ peaks of the M$_1$ phase (indices of the monoclinic structure), which become suppressed in $\sim 140$~fs as shown in the lower panel.
The diffuse intensity integrated between the Bragg peaks, representative of dynamic disorder of the V-V pairs associated with the M$_1$ superstructure and peaks, increases with the same timescale (Fig.~\ref{fig:wall2018}(b), purple symbols). The distribution of diffuse intensity in $\mathbf{Q}$-space resembles the distinct R-phase pattern~\cite{Wall2018}, which originates from transverse acoustic modes that are highly anharmonic, akin to dynamic disorder, and account for large fraction of the lattice entropy\cite{Budai2014metallization}. The transient Rutile diffuse pattern develops extremely fast, indicating that the lattice reaches the disorder in the lattice expected for the Rutile phase in $140$~fs.
These results were confirmed by ab-initio molecular dynamics (AIMD), which showed that the local potential of the V-V pairs (purple double-well trace in Fig.~\ref{fig:wall2018}(c)) is switched quickly to a nearly flat, quartic potential (pink trace in \ref{fig:wall2018}(c)) that allows the V atoms to quickly fill it by randomizing their positions. The absorption of the near-IR pump is modeled by introducing a fictitious electronic temperature $T_\mathrm{el}$. Fig.~\ref{fig:wall2018} shows individual trajectories of this process from the AIMD simulation\cite{Wall2018}. The implications of these results extend beyond VO$_2$: Many ultrafast experiments focus on the behavior of Bragg peaks, while ignoring the role of disorder and fluctuations. This work shows that disorder plays an important role in the lattice of VO$_2$ and ultrafast diffuse scattering not only provides a visualization of the transformation pathway, but it can also yield an alternative view of the thermodynamic transition.

\section{Resonant inelastic x-ray scattering}\label{s:RIXS}
\begin{figure}
\sidecaption
\includegraphics[width=7.5cm]{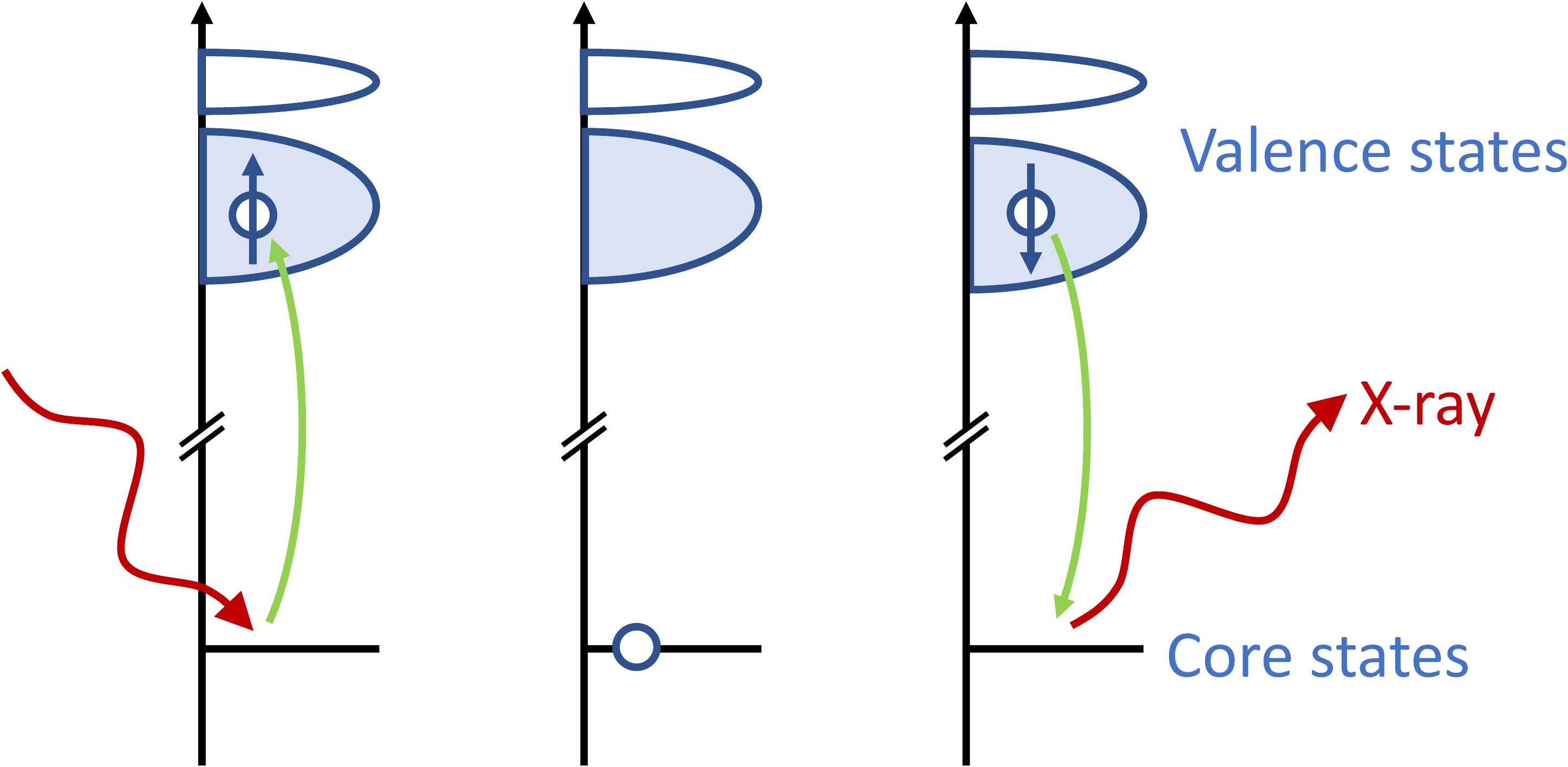}
\caption{A schematic of a direct RIXS process showing the initial state, the intermediate state in which a core electron has been excited, and final states in which a elementary quasiparticle has been created from left to right. The energy and momentum change of the x-ray photon encodes the dispersion of the quasiparticle. }
\label{fig:rixs}
\end{figure}

An important subset of modern x-ray experiments take advantage of x-ray core hole resonances in order to obtain more information than is otherwise possible. Far from resonance x-rays interact with all the electrons in the sample and tend to provide information about structure and structural dynamics such as phonons. Tuning the x-ray energy to a core hole resonance opens routes to couple to other degrees of freedom. \Gls*{RIXS}, as illustrated in Fig.~\ref{fig:rixs}, involves measuring the x-ray energy loss as a function of momentum in order to characterize collective electronic excitations and gain insights into short range correlations and interactions. This can be thought of as the inelastic version of the resonant diffraction experiments described previously (Fig.~\ref{fig:beaud2014} \cite{beaud2014time} and Ref.~\cite{WSLee2012}), where the change in x-ray energy is directly measured in order to infer the properties of excitations beyond simple measurements of the order parameter. Alternatively, it can be conceptualized as an IXS experiment where an x-ray resonance is used to probe spin, charge or orbital excitations as well as than phonons.

The \gls*{RIXS} process is codified in the Kramers-Heisenberg equation, which describes the second order perturbation theory for the interaction between the x-ray photon and a material~\cite{sakurai1967advanced}. The coupling induced in the scattering process can be expressed in terms of electric and magnetic dipole and quadrupole operators (and in special cases to higher order still)~\cite{Joly2012resonant, Ament2011resonant}. These magnetic interactions occur due to the fact that core-hole spin-orbit coupling can, if present, exchange the orbital angular momentum of the photon with the spin in the valence band of the material. For this reason,  standard magnetic dipole resonant scattering involves a rotation of the x-ray polarization -- something that can be exploited in order to identify and isolate magnetic scattering in experiments. Other types of coupling, such as orbital scattering also occur in well-defined channels in the x-ray polarization.

\begin{figure}
\includegraphics[width=\textwidth]{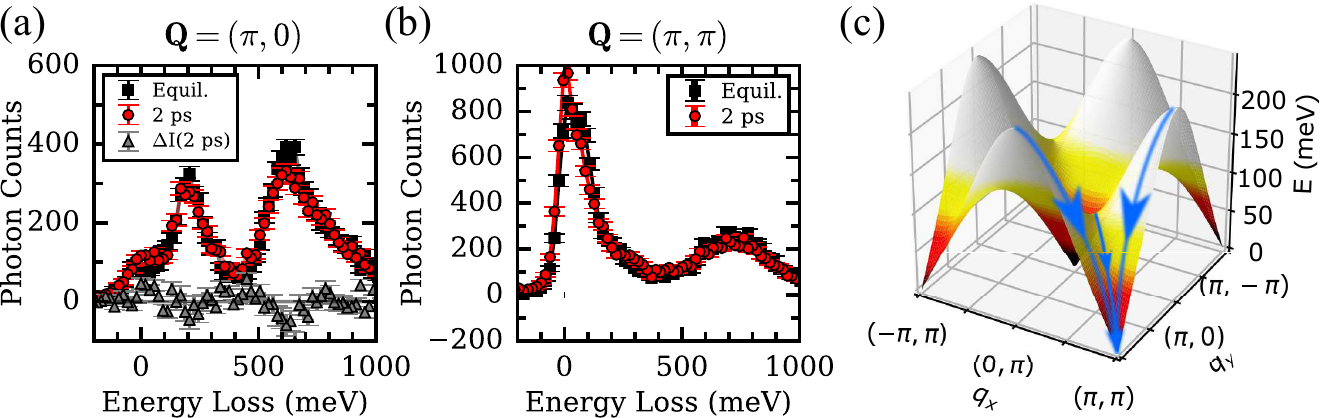}
\caption{Example of ultrafast magnetism probed by RIXS. (a)\&(b) show RIXS spectra of Sr$_2$IrO$_4$ in equilibrium and 2~ps after photo-excitation. Magnetic modes are seen in the 0-200~meV energy window and orbital excitations are seen around 600~meV. The specific $Q$ points measured were (a) $(\pi, 0)$ and (b) $(\pi, \pi)$~\cite{Dean2016ultrafast}. (c) The magnetic dispersion of Sr$_2$IrO$_4$ showing magnons at around 200~meV at $(\pi, 0)$ and almost gapless modes at $(\pi, \pi)$. The blue arrow illustrates the decay of magnons towards the energy-minimum at $(\pi, 0)$ \cite{Mazzone2020trapped}. Adapted by permission from Springer Nature \cite{Dean2016ultrafast}, Copyright (2016).}
\label{fig:Sr2IrO4}
\end{figure}

Syncrotron RIXS has grown dramatically in the past decades driven primarily by improvements in energy resolution that have helped access the low-energy degrees of freedom of correlated materials~\cite{Ament2011resonant, Dean2015insights}. Performing RIXS at XFELs requires the same complicated spectrometer setup to be integrated with an XFEL often together with schemes for photo-exciting the sample. We are, nonetheless, seeing successful examples of time-resolved RIXS, which have targeted charge, orbital and spin degrees of freedom~\cite{Dean2016ultrafast, Cao2019ultrafast, dell2016extreme, Mitrano2019ultrafast, Parchenko2019orbital, Mazzone2020trapped}.

Understanding transient magnetism is an especially important area for ultrafast research~\cite{Kirilyuk2010ultrafast, Gandolfi2017emergent}. Not only is magnetism central to many phenomena including unconventional superconductivity, charge-stripe correlations and quantum spin liquids, but it is also one of the main means for information storage in computers. Recent work by Dean and collaborators performed the first time-resolved RIXS measurement of magnons~\cite{Dean2016ultrafast}. In this study, antiferromagnetic Sr$_2$IrO$_4$ was studied after photoexcitation via a 100~fs, 2~${\mu}$m pulse tuned to excite carriers across the band gap. It was found that a fluence of $6$~mJ/cm$^2$ was sufficient to almost completely suppress long-range magnetic order. The authors proceeded to study the form of the collective magnons in this transient state 2~ps after photoexcitation. As can be seen in Fig.~\ref{fig:Sr2IrO4}, a very similar magnon is observed at the $(\pi, 0)$ reciprocal space point before and after photoexcitation, which suggests that short-range magnetic correlations can continue to exist even while long-range order is strongly suppressed. Since the fluence of the initial excitation process excites a very large fraction of the total magnetic sites, it was  hypothesized that the $(\pi, 0)$ magnon might have been initiallly suppressed, but that it recovers much faster than 2~ps. The low-energy magnon at $(\pi, \pi)$, is modified with respect to the equilibrium configuration at 2~ps (see Fig.~\ref{fig:Sr2IrO4}(b)) and recovers of a few ps timescale. Based on these results, together with resonant elastic scattering data, a picture of a two-stage magnetic recovery was developed in which magnetic correlations are first recovered within the 2D IrO$_2$ planes of Sr$_2$IrO$_4$ and then the registry between planes in achieved in order to return to the ground state. More recently, experiments have been preformed on bilayer antiferromagnetic iridate Sr$_3$Ir$_2$O$_7$. This system features a very large spin gap and a much more gentle dispersion. In this case, magnons at both  $(\pi, 0)$ and $(\pi, \pi)$ reciprocal space points were disturbed in the transient state, which was interpreted in terms of a ``spin-bottleneck'' effect in which the large spin gap trapped the transient magnons over all reciprocal space~\cite{Mazzone2020trapped}.

\begin{figure}
\includegraphics[width=\textwidth]{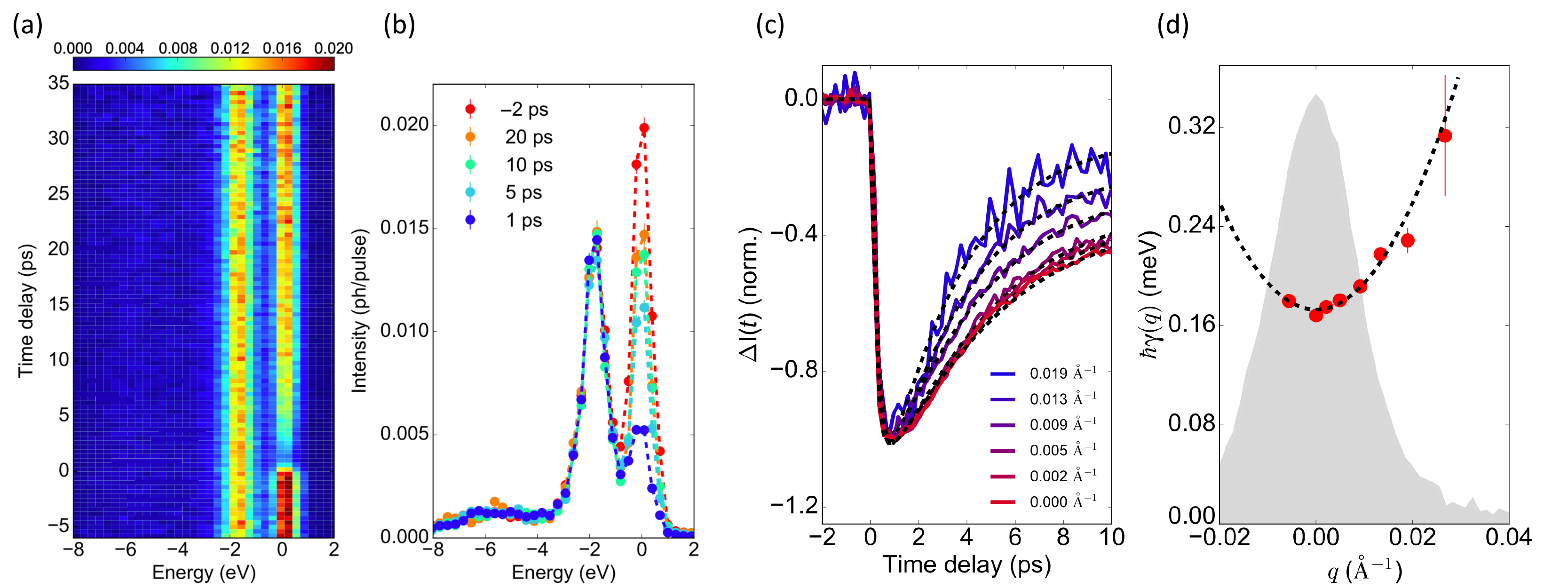}
\caption{RIXS measurements of charge density wave (CDW) dynamics in La$_{2-x}$Ba$_x$CuO$_4$ $x=0.125$~\cite{Mitrano2019ultrafast}. (a)\&(b) show RIXS spectra at $(0.23, 0, 1.5)$ as a function of time delay after a 0.1~mJ/cm$^2$ optical pump pulse represented as either (a) a colormap or (b) over-plotted spectra. The quasi-elastic CDW scattering is seen to first be suppressed and then to recover. (c) time traces of the recovery as a function of Q relative to the peak in the ordering wavevector. (d) The exponential recovery parameter $\hbar \gamma(q)$ as a function the momentum relative to to the peak in the ordering wavevector. The gray shaded area represents the intensity of the CDW in equilibrium. From From~\cite{Mitrano2019ultrafast}. Reprinted with permission from AAAS.}
\label{fig:LBCO}
\end{figure}

RIXS is also known to be highly sensitive to diffuse charge density wave order~\cite{Ament2011resonant, Dean2015insights, miao2017high}. Mitrano et al.\ exploited this effect to analyze the CDW dynamics in cuprate superconductor La$_{2-x}$Ba$_x$CuO$_4$ $x=0.125$ as shown in Fig.~\ref{fig:LBCO}. They observed that the CDW intensity returned to equilibrium without exhibiting any oscillatory behavior. This is consistent with over-damped, diffusive CDW dynamics. The authors outlined a scaling model for the momentum and time dependence of the dynamics~\cite{Mitrano2019ultrafast}. A small, but non-negligible change in the CDW $q$-vector was also observed, which was interpreted in terms of a ``Doppler effect'', in which the pump imparts momentum to the CDW condensate.

\begin{figure}
\includegraphics[width=\textwidth]{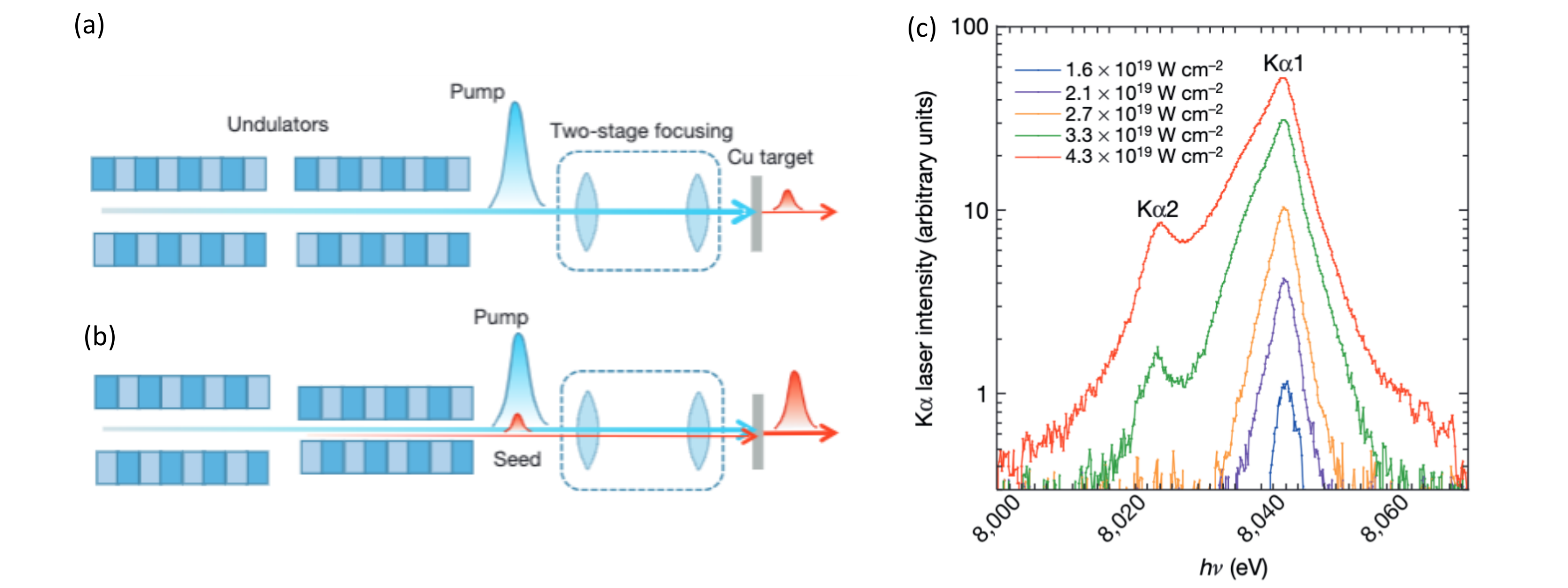}
\caption{Nonlinear processes in the x-ray regime. (a) The experimental setup for demonstrating amplified spontaneous emission, by population inversion for Cu $K_{\alpha}$ emission. X-rays are produced by an undulator and focused to a 120~nm spot on a copper foil. (b) The setup showing amplified emission, in which the pump pulse is accompanied by a seed pulse. (c) Measured Cu $K_\alpha$ emission spectra a different pump intensities. The emitted intensity increases exponentially with incident pump intensity. Adapted by permission from Spinger Nature ~\cite{Yoneda2015atomic}, Copyright (2015).}
\label{fig:xray_lasing}
\end{figure}

To date, the majority of RIXS experiments has been analyzed assuming that the x-rays represent a small perturbation to the material. In this approximation, the spectrum can be interpreted in terms of a single photon interacting with the sample, since the incident photon density is small enough that there is a negligible probability of one photon encountering the effects of other photons. Pioneering work has demonstrated  that FELs now deliver peak photon intensity sufficient to  stimulate  
processes~\cite{Rohringer2012atomic, Beye2013stimulated, Yoneda2015atomic}. Such work has progressed from demonstrations in gases to work on solids, which are the subject of this chapter and the energy employed in the demonstration has extended from extreme ultraviolet energies, through soft x-ray into work with hard x-rays. Figure~\ref{fig:xray_lasing} reproduces work by Yoneda and collaborators, which realizes lasing with the $K_\alpha$ emission line of Cu films~\cite{Yoneda2015atomic}. The authors used a two-stage system to focus the beam from the SACLA XFEL to a 120~nm spot on the sample and monitored the emitted x-rays with a flat silicon crystal spectrometer. Amplified spontaneous emission of x-ray was shown by demonstrating a super-linear dependence of the emitted x-ray intensity as a function of the incident intensity, as plotted in Fig.~\ref{fig:xray_lasing}(c). A notable broadening of the spectrum was reported at high fluences which was attributed to the presence of $3d$ core holes states as well as the target $1s$ core holes excited. Further demonstration of the non-linear interactions was provided by demonstrating seeded emission in which two different color pulses are used, the setup for which is illustrated in Fig.~\ref{fig:xray_lasing}(b). The results (as shown in Ref.~\cite{Yoneda2015atomic}) demonstrate that such a seeded approach can amplify a specific energy emission line, as chosen by providing a narrow-bandwidth seed pulse~\cite{Yoneda2015atomic}. As well as being of intrinsic interest in terms of photonics, such approaches provide new possibilities to perform more efficient measurement that mitigate beam damage or use the enhanced efficiency to access weak multipolar order parameters~\cite{Wang2017on}.

\section{Outlook}

In quantum materials, the complex microscopic interactions result in strong competition between multiple nearly-degenerate ordered phases. Ultrafast pulses provide a strategy for selectively perturbing microscopic interaction parameters to transiently coerce materials into a desired phase\cite{Basov2017towards}. \gls*{XFEL} sources will soon offer high-repetition rate and high-average power\cite{galayda2018,zhu2017sclf}.  The development of cavity-based FEL\cite{marcus2019cavity},  would allow for transform-limited pulses with narrow bandwidth $\sim$~1 meV in the case of an \gls*{XFEL} oscillator (XFELO)~\cite{FELOscillator,kim2009tunable} and  higher peak power with moderate $\sim$100 meV bandwidth  \cite{huang2006fully,marcus2019cavity}in the case of an \gls*{XFEL} regenerative amplifier(RAFEL).  These developments would complement the broadband $\sim$10~eV pulses that allow for near transform limited attosecond pulses~\cite{Huang2017,Duris2020}.  These developments will provide higher-stability and control over the x-ray properties with applications that require either high spectral or temporal resolution,  as well as provide multi-color, multi-pulse capabilities for x-ray pump and x-ray probe measurements of materials dynamics. 

\gls*{XFEL}s will enable comprehensive microscopic characterization of nonequilibrium dynamics in complex materials, by providing unprecedented sensitivity to the collective modes. 
For example, high repetition rate sources will  facilitate visualization of nanoscale heat conduction with potential prospects for microscopic control of energy flow. Novel approaches have emerged for control of sound and heat exploiting (coherent) interferences of waves within structures comparable to the wavelength \cite{maldovan}.  
The impact of new \gls*{XFEL} sources for probing nanoscale heat transport will be in probing the dynamics of high wavevector phonons with nm wavelenths at femtosecond timescales, yielding a microscopic view of wave propagation and heat conduction.

In addition, the combination of high repetition-rate and tunable photon energy is well-matched to studying novel light-driven quantum states through resonant scattering, while multi-color, multi-pulse and attosecond operation will enable a much deeper understanding of nonlinear light-matter interactions in quantum materials.  Time-resolved inelastic scattering measurements will be possible in a variety of sample environments such as inside a diamond anvil cell.
This will enable detailed measurements of quantum critical fluctuations of charge and spin degrees of freedom,  their associated collective modes and their couplings, while tuning continuously through a quantum critical point. Such unique capabilities will improve our understanding of the connection between superconductivity and the fluctuations in other order parameters~\cite{Joe2014emergence,Feng2015itinerant}. These are just some examples of how the revolutionary properties of x-ray free electron lasers are likely to make an lasting  impact on important problems in materials science and condensed matter physics.

This work was supported by the U.S.\ Department of Energy, Office of Science, Office of Basic Energy Sciences through the Division of Materials Sciences and Engineering under Contract Nos.~DE-AC02-76SF00515 (MT and DAR) and~DE-de-sc0012704 (MPMD). 

\bibliographystyle{spphys}

\begin{thebibliography}{100}
\providecommand{\url}[1]{{#1}}
\providecommand{\urlprefix}{URL }
\expandafter\ifx\csname urlstyle\endcsname\relax
  \providecommand{\doi}[1]{DOI \discretionary{}{}{}#1}\else
  \providecommand{\doi}{DOI \discretionary{}{}{}\begingroup
  \urlstyle{rm}\Url}\fi

\bibitem{krisch2007inelastic}
M.~Krisch, F.~Sette, in \emph{Light Scattering in Solid IX} (Springer, 2007),
  pp. 317--370

\bibitem{Baron2015}
A.Q.R. Baron, in \emph{Synchrotron Light Sources and Free-Electron Lasers:
  Accelerator Physics, Instrumentation and Science Applications}, ed. by
  E.~Jaeschke, S.~Khan, J.R. Schneider, J.B. Hastings (Springer International
  Publishing, 2015), pp. 1--68

\bibitem{Ament2011resonant}
L.J.P. Ament, M.~van Veenendaal, T.P. Devereaux, J.P. Hill, J.~van~den Brink,
  Rev. Mod. Phys. \textbf{83}, 705 (2011)

\bibitem{Dean2015insights}
M.~Dean, Journal of Magnetism and Magnetic Materials \textbf{376}, 3  (2015)

\bibitem{wall2016recent}
S.~Wall, M.~Trigo, Synchrotron Radiation News \textbf{29}(5), 13 (2016)

\bibitem{Lindenberg2017}
A.M. Lindenberg, S.L. Johnson, D.A. Reis, Annual Review of Materials Research
  \textbf{47}(1), 425 (2017)

\bibitem{Buzzi2018probing}
M.~Buzzi, M.~F{\"o}rst, R.~Mankowsky, A.~Cavalleri, Nature Reviews Materials
  \textbf{3}(9), 299 (2018)

\bibitem{Dunne2018x-ray}
M.~Dunne, Nature Reviews Materials \textbf{3}(9), 290 (2018)

\bibitem{Cao2019ultrafast}
Y.~Cao, D.G. Mazzone, D.~Meyers, J.P. Hill, X.~Liu, S.~Wall, M.P.M. Dean,
  Philosophical Transactions of the Royal Society A: Mathematical, Physical and
  Engineering Sciences \textbf{377}(2145), 20170480 (2019)

\bibitem{Emma2010first}
P.~Emma, R.~Akre, J.~Arthur, R.~Bionta, C.~Bostedt, J.~Bozek, A.~Brachmann,
  P.~Bucksbaum, R.~Coffee, F.J. Decker, et~al., nature photonics \textbf{4}(9),
  641 (2010)

\bibitem{Kondratenko1980generating}
A.~Kondratenko, E.~Saldin, Part. Accel. \textbf{10}, 207 (1980)

\bibitem{Bonifacio1984collective}
R.~Bonifacio, C.~Pellegrini, L.~Narducci, in \emph{AIP Conference Proceedings},
  vol. 118 (American Institute of Physics, 1984), vol. 118, pp. 236--259

\bibitem{Bonifacio:1984aa}
R.~Bonifacio, C.~Pellegrini, L.M. Narducci, Optics Communications
  \textbf{50}(6), 373 (1984)

\bibitem{Ishikawa2012compact}
T.~Ishikawa, H.~Aoyagi, T.~Asaka, Y.~Asano, N.~Azumi, T.~Bizen, H.~Ego,
  K.~Fukami, T.~Fukui, Y.~Furukawa, et~al., nature photonics \textbf{6}(8), 540
  (2012)

\bibitem{Abeghyan2019}
S.~Abeghyan, M.~Bagha-Shanjani, G.~Chen, U.~Englisch, S.~Karabekyan, Y.~Li,
  F.~Preisskorn, F.~Wolff-Fabris, M.~Wuenschel, M.~Yakopov, J.~Pflueger,
  Journal of Synchrotron Radiation \textbf{26}(2), 302 (2019)

\bibitem{Kang2017hard}
H.S. Kang, C.K. Min, H.~Heo, C.~Kim, H.~Yang, G.~Kim, I.~Nam, S.Y. Baek, H.J.
  Choi, G.~Mun, et~al., Nature Photonics \textbf{11}(11), 708 (2017)

\bibitem{milne2017swissfel}
C.J. Milne, T.~Schietinger, M.~Aiba, A.~Alarcon, J.~Alex, A.~Anghel, V.~Arsov,
  C.~Beard, P.~Beaud, S.~Bettoni, et~al., Applied Sciences \textbf{7}(7), 720
  (2017)

\bibitem{Amann2012demonstration}
J.~Amann, W.~Berg, V.~Blank, F.J. Decker, Y.~Ding, P.~Emma, Y.~Feng, J.~Frisch,
  D.~Fritz, J.~Hastings, et~al., Nature photonics \textbf{6}(10), 693 (2012)

\bibitem{Inoue2019}
I.~Inoue, T.~Osaka, T.~Hara, T.~Tanaka, T.~Inagaki, T.~Fukui, S.~Goto,
  Y.~Inubushi, H.~Kimura, R.~Kinjo, H.~Ohashi, K.~Togawa, K.~Tono, M.~Yamaga,
  H.~Tanaka, T.~Ishikawa, M.~Yabashi, Nature Photonics \textbf{13}(5), 319
  (2019)

\bibitem{Huang2017}
S.~Huang, Y.~Ding, Y.~Feng, E.~Hemsing, Z.~Huang, J.~Krzywinski, A.A. Lutman,
  A.~Marinelli, T.J. Maxwell, D.~Zhu, Phys. Rev. Lett. \textbf{119}, 154801
  (2017)

\bibitem{Duris2020}
J.~Duris, S.~Li, T.~Driver, E.G. Champenois, J.P. MacArthur, A.A. Lutman,
  Z.~Zhang, P.~Rosenberger, J.W. Aldrich, R.~Coffee, G.~Coslovich, F.J. Decker,
  J.M. Glownia, G.~Hartmann, W.~Helml, A.~Kamalov, J.~Knurr, J.~Krzywinski,
  M.F. Lin, J.P. Marangos, M.~Nantel, A.~Natan, J.T. O'Neal, N.~Shivaram,
  P.~Walter, A.L. Wang, J.J. Welch, T.J.A. Wolf, J.Z. Xu, M.F. Kling, P.H.
  Bucksbaum, A.~Zholents, Z.~Huang, J.P. Cryan, A.~Marinelli, Nature Photonics
  \textbf{14}(1), 30 (2020)

\bibitem{Tono_2013}
K.~Tono, T.~Togashi, Y.~Inubushi, T.~Sato, T.~Katayama, K.~Ogawa, H.~Ohashi,
  H.~Kimura, S.~Takahashi, K.~Takeshita, H.~Tomizawa, S.~Goto, T.~Ishikawa,
  M.~Yabashi, New Journal of Physics \textbf{15}(8), 083035 (2013)

\bibitem{Hara:2013aa}
T.~Hara, Y.~Inubushi, T.~Katayama, T.~Sato, H.~Tanaka, T.~Tanaka, T.~Togashi,
  K.~Togawa, K.~Tono, M.~Yabashi, T.~Ishikawa, Nature Communications
  \textbf{4}(1), 2919 (2013)

\bibitem{galayda2018}
J.~Galayda, et~al., in \emph{9th Int. Particle Accelerator Conf.(IPAC'18),
  Vancouver, BC, Canada, April 29-May 4, 2018} (2018), pp. 18--23

\bibitem{zhu2017sclf}
Z.~Zhu, Z.~Zhao, D.~Wang, Z.~Liu, R.~Li, L.~Yin, Z.~Yang, Proceedings of the
  FEL2017, Santa Fe, NM, USA pp. 20--25 (2017)

\bibitem{Huang2007review}
Z.~Huang, K.J. Kim, Phys. Rev. ST Accel. Beams \textbf{10}, 034801 (2007)

\bibitem{saldin2010statistical}
E.L. Saldin, E.A. Schneidmiller, M.~Yurkov, New Journal of Physics
  \textbf{12}(3), 035010 (2010)

\bibitem{McNeil2010}
B.W.J. McNeil, N.R. Thompson, Nature Photonics \textbf{4}(12), 814 (2010)

\bibitem{Pellegrini2016pellegrini}
C.~Pellegrini, A.~Marinelli, S.~Reiche, Rev. Mod. Phys. \textbf{88}, 015006
  (2016).
\newblock \doi{10.1103/RevModPhys.88.015006}

\bibitem{seddon2017}
E.~Seddon, J.~Clarke, D.~Dunning, C.~Masciovecchio, C.~Milne, F.~Parmigiani,
  D.~Rugg, J.~Spence, N.~Thompson, K.~Ueda, et~al., Reports on Progress in
  Physics \textbf{80}(11), 115901 (2017)

\bibitem{Nugent2010coherent}
K.A. Nugent, Advances in Physics \textbf{59}(1), 1 (2010)

\bibitem{Shpyrko2014xray}
O.G. Shpyrko, Journal of Synchrotron Radiation \textbf{21}(5), 1057 (2014)

\bibitem{Glover2012}
T.E. Glover, D.M. Fritz, M.~Cammarata, T.K. Allison, S.~Coh, J.M. Feldkamp,
  H.~Lemke, D.~Zhu, Y.~Feng, R.N. Coffee, M.~Fuchs, S.~Ghimire, J.~Chen,
  S.~Shwartz, D.A. Reis, S.E. Harris, J.B. Hastings, Nature \textbf{488}(7413),
  603 (2012)

\bibitem{Shwartz2014}
S.~Shwartz, M.~Fuchs, J.B. Hastings, Y.~Inubushi, T.~Ishikawa, T.~Katayama,
  D.A. Reis, T.~Sato, K.~Tono, M.~Yabashi, S.~Yudovich, S.E. Harris, Phys. Rev.
  Lett. \textbf{112}, 163901 (2014)

\bibitem{Tamasaku:2014qf}
K.~Tamasaku, E.~Shigemasa, Y.~Inubushi, T.~Katayama, K.~Sawada, H.~Yumoto,
  H.~Ohashi, H.~Mimura, M.~Yabashi, K.~Yamauchi, T.~Ishikawa, Nat Photon
  \textbf{8}(4), 313 (2014)

\bibitem{Fuchs2015}
M.~Fuchs, M.~Trigo, J.~Chen, S.~Ghimire, S.~Shwartz, M.~Kozina, M.~Jiang,
  T.~Henighan, C.~Bray, G.~Ndabashimiye, P.H. Bucksbaum, Y.~Feng, S.~Herrmann,
  G.A. Carini, J.~Pines, P.~Hart, C.~Kenney, S.~Guillet, S.~Boutet, G.J.
  Williams, M.~Messerschmidt, M.M. Seibert, S.~Moeller, J.B. Hastings, D.A.
  Reis, Nat Phys \textbf{11}(11), 964 (2015)

\bibitem{Ghimire2016}
S.~Ghimire, M.~Fuchs, J.~Hastings, S.C. Herrmann, Y.~Inubushi, J.~Pines,
  S.~Shwartz, M.~Yabashi, D.A. Reis, Phys. Rev. A \textbf{94}, 043418 (2016)

\bibitem{Fuchs2018Young}
M.~Fuchs, D.A. Reis, in \emph{Young, L. et al., Roadmap of ultrafast x-ray
  atomic and molecular physics}, vol.~51 (Journal of Physics B: Atomic,
  Molecular and Optical Physics, 2018), p. 032003

\bibitem{Murnane1991ultrafast}
M.M. Murnane, H.C. Kapteyn, M.D. Rosen, R.W. Falcone, Science
  \textbf{251}(4993), 531 (1991)

\bibitem{Sokolowski_Tinten_2004}
K.~Sokolowski-Tinten, D.~von~der Linde, Journal of Physics: Condensed Matter
  \textbf{16}(49), R1517 (2004)

\bibitem{Schoenlein2000generation}
R.~Schoenlein, S.~Chattopadhyay, H.~Chong, T.~Glover, P.~Heimann, C.~Shank,
  A.~Zholents, M.~Zolotorev, Science \textbf{287}(5461), 2237 (2000)

\bibitem{bentson2003}
L.~Bentson, P.~Bolton, E.~Bong, P.~Emma, J.~Galayda, J.~Hastings, P.~Krejcik,
  C.~Rago, J.~Rifkin, C.~Spencer, Nuclear Instruments and Methods in Physics
  Research Section A: Accelerators, Spectrometers, Detectors and Associated
  Equipment \textbf{507}(1-2), 205 (2003)

\bibitem{Lindenberg2005}
A.M. Lindenberg, J.~Larsson, K.~Sokolowski-Tinten, K.J. Gaffney, C.~Blome,
  O.~Synnergren, J.~Sheppard, C.~Caleman, A.G. {MacPhee}, D.~Weinstein, D.P.
  Lowney, T.K. Allison, T.~Matthews, R.W. Falcone, A.L. Cavalieri, D.M. Fritz,
  S.H. Lee, P.H. Bucksbaum, D.A. Reis, J.~Rudati, P.H. Fuoss, C.C. Kao, D.P.
  Siddons, R.~Pahl, J.~Als-Nielson, S.~Duerster, I.~R., H.~Schlarb,
  H.~Schulte-Schrepping, T.~Tschentscher, J.~Schneider, D.~von~der Linde,
  O.~Hignette, F.~Sette, H.N. Chapman, R.W. Lee, T.N. Hansen, S.~Techert, J.S.
  Wark, M.~Bergh, G.~Huldt, D.~van~der Spoel, T.~N., J.~Hajdu, R.A. Akre,
  E.~Bong, P.~Krejcik, J.~Arthur, S.~Brennan, L.~K., J.B. Hastings, Science
  \textbf{308}, 392 (2005)

\bibitem{Cavalieri2005}
A.L. Cavalieri, D.M. Fritz, S.H. Lee, P.H. Bucksbaum, D.A. Reis, J.~Rudati,
  D.M. Mills, P.H. Fuoss, G.B. Stephenson, C.C. Kao, D.P. Siddons, D.P. Lowney,
  A.G. {MacPhee}, D.~Weinstein, R.W. Falcone, R.~Pahl, J.~Als-Nielsen,
  C.~Blome, S.~D{\"u}rsterer, R.~Ischebeck, H.~Schlarb, H.~Schulte-Schrepping,
  T.~Tschentscher, J.~Schneider, O.~Hignette, F.~Sette, k.~Sokolowski-Tinten,
  H.N. Chapman, R.W. Lee, T.N. Hansen, O.~Synnergren, J.~Larsson, S.~Techert,
  J.~Sheppard, J.S. Wark, M.~Bergh, C.~Caleman, G.~Huldt, D.~van~der Spoel,
  N.~Timneanu, J.~Hajdu, R.A. Akre, E.~Bong, P.~Emma, P.~Krejcik, J.~Arthur,
  S.~Brennan, K.J. Gaffney, A.M. Lindenberg, K.~Luening, J.B. Hastings, Phys.
  Rev. Lett. \textbf{94}, 114801 (2005)

\bibitem{Rousse2001rmp}
A.~Rousse, C.~Rischel, J.C. Gauthier, Rev. Mod. Phys. \textbf{73}, 17 (2001)

\bibitem{Cavalleri2002Review}
A.~Cavalleri, C.~Blome, P.~Forget, J.~Kieffer, S.~Magnan, C.~Siders,
  K.~Sokolowski-Tinten, J.~Squier, C.~T{\'O}th, D.V. Linde, Phase Transitions
  \textbf{75}(7-8), 769 (2002)

\bibitem{Reis2006ultrafast}
D.A. Reis, A.M. Lindenberg, in \emph{Light Scattering in Solid IX} (Springer,
  2006), pp. 371--422

\bibitem{bargheer2006recent}
M.~Bargheer, N.~Zhavoronkov, M.~Woerner, T.~Elsaesser, Chemphyschem: a European
  journal of chemical physics and physical chemistry \textbf{7}(4), 783 (2006)

\bibitem{Bionta2011spectral}
M.~Bionta, H.~Lemke, J.~Cryan, J.~Glownia, C.~Bostedt, M.~Cammarata,
  J.~Castagna, Y.~Ding, D.~Fritz, A.~Fry, J.~Krzywinski, M.~Messerschmidt,
  S.~Schorb, M.~Swiggers, R.~Coffee, Optics Express \textbf{19}(22), 21855
  (2011)

\bibitem{chollet2015}
M.~Chollet, R.~Alonso-Mori, M.~Cammarata, D.~Damiani, J.~Defever, J.T. Delor,
  Y.~Feng, J.M. Glownia, J.B. Langton, S.~Nelson, K.~Ramsey, A.~Robert,
  M.~Sikorski, S.~Song, D.~Stefanescu, V.~Srinivasan, D.~Zhu, H.T. Lemke, D.M.
  Fritz, Journal of Synchrotron Radiation \textbf{22}(3), 503 (2015)

\bibitem{dhar1994timeresolved}
L.~Dhar, J.A. Rogers, K.A. Nelson, Chemical Reviews \textbf{94}(1), 157 (1994)

\bibitem{merlin1997generating}
R.~Merlin, Solid State Communications \textbf{102}(2--3), 207  (1997)

\bibitem{Singer2016}
A.~Singer, S.K.K. Patel, R.~Kukreja, V.~Uhl\'{\i}\ifmmode~\check{r}\else
  \v{r}\fi{}, J.~Wingert, S.~Festersen, D.~Zhu, J.M. Glownia, H.T. Lemke,
  S.~Nelson, M.~Kozina, K.~Rossnagel, M.~Bauer, B.M. Murphy, O.M. Magnussen,
  E.E. Fullerton, O.G. Shpyrko, Phys. Rev. Lett. \textbf{117}, 056401 (2016)

\bibitem{kozina2017heterodyne}
M.~Kozina, M.~Trigo, M.~Chollet, J.N. Clark, J.M. Glownia, A.C. Gossard,
  T.~Henighan, M.P. Jiang, H.~Lu, A.~Majumdar, D.~Zhu, D.A. Reis, Structural
  Dynamics \textbf{4}(5), 054305 (2017)

\bibitem{Zeiger1992theory}
H.J. Zeiger, J.~Vidal, T.K. Cheng, E.P. Ippen, G.~Dresselhaus, M.S.
  Dresselhaus, Phys. Rev. B \textbf{45}, 768 (1992)

\bibitem{Gerber2017femtosecond}
S.~Gerber, S.L. Yang, D.~Zhu, H.~Soifer, J.A. Sobota, S.~Rebec, J.J. Lee,
  T.~Jia, B.~Moritz, C.~Jia, A.~Gauthier, Y.~Li, D.~Leuenberger, Y.~Zhang,
  L.~Chaix, W.~Li, H.~Jang, J.S. Lee, M.~Yi, G.L. Dakovski, S.~Song, J.M.
  Glownia, S.~Nelson, K.W. Kim, Y.D. Chuang, Z.~Hussain, R.G. Moore, T.P.
  Devereaux, W.S. Lee, P.S. Kirchmann, Z.X. Shen, Science \textbf{357}(6346),
  71 (2017)

\bibitem{Trigo2013Fourier}
M.~Trigo, M.~Fuchs, J.~Chen, M.P. Jiang, M.~Cammarata, S.~Fahy, D.M. Fritz,
  K.~Gaffney, S.~Ghimire, A.~Higginbotham, S.L. Johnson, M.E. Kozina,
  J.~Larsson, H.~Lemke, A.M. Lindenberg, G.~Ndabashimiye, F.~Quirin,
  K.~Sokolowski-Tinten, C.~Uher, G.~Wang, J.S. Wark, D.~Zhu, D.A. Reis, Nature
  Physics \textbf{9}(12), 790 (2013)

\bibitem{Warren1969Xray}
B.E. Warren, \emph{X-Ray Diffraction} (Dover, New York, 1969)

\bibitem{Zhu2015Phonon}
D.~Zhu, A.~Robert, T.~Henighan, H.T. Lemke, M.~Chollet, J.M. Glownia, D.A.
  Reis, M.~Trigo, Phys. Rev. B \textbf{92} (2015)

\bibitem{Imada1998metal}
M.~Imada, A.~Fujimori, Y.~Tokura, Rev. Mod. Phys. \textbf{70}, 1039 (1998).
\newblock \doi{10.1103/RevModPhys.70.1039}.
\newblock \urlprefix\url{https://link.aps.org/doi/10.1103/RevModPhys.70.1039}

\bibitem{Foerst2011nonlinear}
M.~F{\"o}rst, C.~Manzoni, S.~Kaiser, Y.~Tomioka, Y.~Tokura, R.~Merlin,
  A.~Cavalleri, Nature Physics \textbf{7}(11), 854 (2011)

\bibitem{Mankowsky2014}
R.~Mankowsky, A.~Subedi, M.~F{\"o}rst, S.O. Mariager, M.~Chollet, H.T. Lemke,
  J.S. Robinson, J.M. Glownia, M.P. Minitti, A.~Frano, M.~Fechner, N.A.
  Spaldin, T.~Loew, B.~Keimer, A.~Georges, A.~Cavalleri, Nature
  \textbf{516}(7529), 71 (2014)

\bibitem{Kozina2019}
M.~Kozina, M.~Fechner, P.~Marsik, T.~van Driel, J.M. Glownia, C.~Bernhard,
  M.~Radovic, D.~Zhu, S.~Bonetti, U.~Staub, M.C. Hoffmann, Nature Physics
  \textbf{15}(4), 387 (2019)

\bibitem{Juraschek2018Sumfrequency}
D.M. Juraschek, S.F. Maehrlein, Phys. Rev. B p. 174302 (2018)

\bibitem{Fausti2011light}
D.~Fausti, R.~Tobey, N.~Dean, S.~Kaiser, A.~Dienst, M.C. Hoffmann, S.~Pyon,
  T.~Takayama, H.~Takagi, A.~Cavalleri, science \textbf{331}(6014), 189 (2011)

\bibitem{Orbach}
R.~Orbach, Phys. Rev. Lett. \textbf{16}, 15 (1966)

\bibitem{Klemens}
P.G. Klemens, Phys. Rev. \textbf{148}, 845 (1966)

\bibitem{Debernardi1995}
A.~Debernardi, S.~Baroni, E.~Molinari, Phys. Rev. Lett. \textbf{75}, 1819
  (1995)

\bibitem{Debernardi1998}
A.~Debernardi, Phys. Rev. B \textbf{57}, 12847 (1998)

\bibitem{Togo2015Distributions}
A.~Togo, L.~Chaput, I.~Tanaka, Phys. Rev. B \textbf{91}, 094306 (2015)

\bibitem{Teitelbaum2018}
S.W. Teitelbaum, T.~Henighan, Y.~Huang, H.~Liu, M.P. Jiang, D.~Zhu, M.~Chollet,
  T.~Sato, E.D. Murray, S.~Fahy, S.~O'Mahony, T.P. Bailey, C.~Uher, M.~Trigo,
  D.A. Reis, Phys. Rev. Lett. \textbf{121}, 125901 (2018)

\bibitem{Cheng1990}
T.K. Cheng, S.D. Brorson, A.S. Kazeroonian, J.S. Moodera, G.~Dresselhaus, M.S.
  Dresselhaus, E.P. Ippen, Appl. Phys. Lett. \textbf{57}, 1004 (1990)

\bibitem{deCamp2001PRB}
M.F. DeCamp, D.A. Reis, P.H. Bucksbaum, R.~Merlin, Phys. Rev. B. \textbf{64},
  092301 (2001)

\bibitem{hase1996}
M.~Hase, K.~Mizoguchi, H.~Harima, S.~Nakashima, M.~Tani, K.~Sakai, M.~Hangyo,
  Appl. Phys. Lett. \textbf{69}(17), 2474 (1996)

\bibitem{hase2002}
M.~Hase, M.~Kitajima, S.~Nakashima, K.~Mizoguchi, Phys. Rev. Lett.
  \textbf{88}(6), 067401 (2002)

\bibitem{Murray2005}
E.D. Murray, D.M. Fritz, J.K. Wahlstrand, S.~Fahy, D.A. Reis, Phys. Rev. B
  \textbf{72}(6), 060301 (2005)

\bibitem{teitelbaum2018real}
S.W. Teitelbaum, T.~Shin, J.W. Wolfson, Y.H. Cheng, I.J. Porter, M.~Kandyla,
  K.A. Nelson, Physical Review X \textbf{8}(3), 031081 (2018)

\bibitem{Sokolowski-Tinten2003}
K.~Sokolowski-Tinten, C.~Blome, J.~Blums, A.~Cavalleri, C.~Dietrich,
  A.~Tarasevitch, I.~Uschmann, E.~F{\"o}rster, M.~Kamller, M.~Horn~von Hoegen,
  D.~von~der Linde, Nature \textbf{422}, 287 (2003)

\bibitem{Fritz2007}
D.M. Fritz, D.A. Reis, B.~Adams, R.A. Akre, J.~Arthur, C.~Blome, P.H.
  Bucksbaum, A.L. Cavalieri, S.~Engemann, S.~Fahy, R.W. Falcone, P.H. Fuoss,
  K.J. Gaffney, M.J. George, J.~Hajdu, M.P. Hertlein, P.B. Hillyard,
  M.~Horn-von Hoegen, M.~Kammler, J.~Kaspar, R.~Kienberger, P.~Krejcik, S.H.
  Lee, A.M. Lindenberg, B.~McFarland, D.~Meyer, T.~Montagne, {\'E}.D. Murray,
  A.J. Nelson, M.~Nicoul, R.~Pahl, J.~Rudati, H.~Schlarb, D.P. Siddons,
  K.~Sokolowski-Tinten, T.~Tschentscher, D.~von~der Linde, J.B. Hastings,
  Science \textbf{315}(5812), 633 (2007)

\bibitem{Johnson2008nanoscale}
S.L. Johnson, P.~Beaud, C.J. Milne, F.S. Krasniqi, E.S. Zijlstra, M.E. Garcia,
  M.~Kaiser, D.~Grolimund, R.~Abela, G.~Ingold, Phys. Rev. Lett. \textbf{100},
  155501 (2008)

\bibitem{Fahy2016}
S.~Fahy, E.D. Murray, D.A. Reis, Phys. Rev. B \textbf{93}, 134308 (2016)

\bibitem{Trigo2019Coherent}
M.~Trigo, P.~Giraldo-Gallo, M.E. Kozina, T.~Henighan, M.P. Jiang, H.~Liu, J.N.
  Clark, M.~Chollet, J.M. Glownia, D.~Zhu, T.~Katayama, D.~Leuenberger, P.S.
  Kirchmann, I.R. Fisher, Z.X. Shen, D.A. Reis, Phys. Rev. B \textbf{99},
  104111 (2019)

\bibitem{ru2008}
N.~Ru, C.L. Condron, G.Y. Margulis, K.Y. Shin, J.~Laverock, S.B. Dugdale, M.F.
  Toney, I.R. Fisher, Phys. Rev. B \textbf{77}, 035114 (2008)

\bibitem{goldenfeld1992lectures}
N.~Goldenfeld, \emph{Lectures on phase transitions and the renormalization
  group}.
\newblock Frontiers in physics (Addison-Wesley, Advanced Book Program, 1992)

\bibitem{Schaefer2014Collective}
H.~Schaefer, V.V. Kabanov, J.~Demsar, Phys. Rev. B \textbf{89}, 045106 (2014)

\bibitem{Huber2014Coherent}
T.~Huber, S.O. Mariager, A.~Ferrer, H.~Sch\"afer, J.A. Johnson, S.~Gr\"ubel,
  A.~L\"ubcke, L.~Huber, T.~Kubacka, C.~Dornes, C.~Laulhe, S.~Ravy, G.~Ingold,
  P.~Beaud, J.~Demsar, S.L. Johnson, Phys. Rev. Lett. \textbf{113}, 026401
  (2014)

\bibitem{beaud2014time}
P.~Beaud, A.~Caviezel, S.~Mariager, L.~Rettig, G.~Ingold, C.~Dornes, S.~Huang,
  J.~Johnson, M.~Radovic, T.~Huber, et~al., Nature materials \textbf{13}(10),
  923 (2014)

\bibitem{Chuang2013RealTime}
Y.D. Chuang, W.S. Lee, Y.F. Kung, A.P. Sorini, B.~Moritz, R.G. Moore,
  L.~Patthey, M.~Trigo, D.H. Lu, P.S. Kirchmann, M.~Yi, O.~Krupin, M.~Langner,
  Y.~Zhu, S.Y. Zhou, D.A. Reis, N.~Huse, J.S. Robinson, R.A. Kaindl, R.W.
  Schoenlein, S.L. Johnson, M.~F\"orst, D.~Doering, P.~Denes, W.F. Schlotter,
  J.J. Turner, T.~Sasagawa, Z.~Hussain, Z.X. Shen, T.P. Devereaux, Phys. Rev.
  Lett. \textbf{110}, 127404 (2013)

\bibitem{Chuang2013real}
Y.D. Chuang, W.S. Lee, Y.F. Kung, A.P. Sorini, B.~Moritz, R.G. Moore,
  L.~Patthey, M.~Trigo, D.H. Lu, P.S. Kirchmann, M.~Yi, O.~Krupin, M.~Langner,
  Y.~Zhu, S.Y. Zhou, D.A. Reis, N.~Huse, J.S. Robinson, R.A. Kaindl, R.W.
  Schoenlein, S.L. Johnson, M.~F\"orst, D.~Doering, P.~Denes, W.F. Schlotter,
  J.J. Turner, T.~Sasagawa, Z.~Hussain, Z.X. Shen, T.P. Devereaux, Phys. Rev.
  Lett. \textbf{110}, 127404 (2013)

\bibitem{WSLee2012}
W.S. Lee, Y.D. Chuang, R.G. Moore, Y.~Zhu, L.~Patthey, M.~Trigo, D.H. Lu, P.S.
  Kirchmann, O.~Krupin, M.~Yi, M.~Langner, N.~Huse, J.S. Robinson, Y.~Chen,
  S.Y. Zhou, G.~Coslovich, B.~Huber, D.A. Reis, R.A. Kaindl, R.W. Schoenlein,
  D.~Doering, P.~Denes, W.F. Schlotter, J.J. Turner, S.L. Johnson,
  M.~F{\"o}rst, T.~Sasagawa, Y.F. Kung, A.P. Sorini, A.F. Kemper, B.~Moritz,
  T.P. Devereaux, D.H. Lee, Z.X. Shen, Z.~Hussain, Nature Communications
  \textbf{3}(1), 838 (2012)

\bibitem{Wall2018}
S.~Wall, S.~Yang, L.~Vidas, M.~Chollet, J.M. Glownia, M.~Kozina, T.~Katayama,
  T.~Henighan, M.~Jiang, T.A. Miller, D.A. Reis, L.A. Boatner, O.~Delaire,
  M.~Trigo, Science \textbf{362}(6414), 572 (2018)

\bibitem{Morin1959Oxides}
F.J. Morin, Phys. Rev. Lett. \textbf{3}, 34 (1959)

\bibitem{Cavalleri2001Femtosecond}
A.~Cavalleri, C.~T\'oth, C.W. Siders, J.A. Squier, F.~R\'aksi, P.~Forget, J.C.
  Kieffer, Phys. Rev. Lett. \textbf{87}, 237401 (2001)

\bibitem{Budai2014metallization}
J.D. Budai, J.~Hong, M.E. Manley, E.D. Specht, C.W. Li, J.Z. Tischler, D.L.
  Abernathy, A.H. Said, B.M. Leu, L.A. Boatner, et~al., Nature
  \textbf{515}(7528), 535 (2014)

\bibitem{sakurai1967advanced}
J.~Sakurai, \emph{Advanced Quantum Mechanics}.
\newblock Always learning (Pearson Education, Incorporated, 1967)

\bibitem{Joly2012resonant}
Y.~Joly, S.D. Matteo, O.~Bun{\u{a}}u, The European Physical Journal Special
  Topics \textbf{208}(1), 21 (2012)

\bibitem{Dean2016ultrafast}
M.P.M. Dean, Y.~Cao, X.~Liu, S.~Wall, D.~Zhu, R.~Mankowsky, V.~Thampy, X.M.
  Chen, J.G. Vale, D.~Casa, J.~Kim, A.H. Said, P.~Juhas, R.~Alonso-Mori, J.M.
  Glownia, A.~Robert, J.~Robinson, M.~Sikorski, S.~Song, M.~Kozina, H.~Lemke,
  L.~Patthey, S.~Owada, T.~Katayama, M.~Yabashi, Y.~Tanaka, T.~Togashi, J.~Liu,
  C.~Rayan~Serrao, B.J. Kim, L.~Huber, C.L. Chang, D.F. McMorrow, M.~F{\"o}rst,
  J.P. Hill, Nature Materials \textbf{15}, 601 (2016)

\bibitem{Mazzone2020trapped}
D.G. {Mazzone}, D.~{Meyers}, Y.~{Cao}, J.G. {Vale}, C.D. {Dashwood}, Y.~{Shi},
  A.J.A. {James}, N.J. {Robinson}, J.Q. {Lin}, V.~{Thampy}, Y.~{Tanaka}, A.S.
  {Johnson}, H.~{Miao}, R.~{Wang}, T.A. {Assefa}, J.~{Kim}, D.~{Casa},
  R.~{Mankowsky}, D.~{Zhu}, R.~{Alonso-Mori}, S.~{Song}, H.~{Yavas},
  T.~{Katayama}, M.~{Yabashi}, Y.K.S. {Owada}, J.~{Liu}, J.~{Yang}, R.M.
  {Konik}, I.K. {Robinson}, J.P. {Hill}, D.F. {McMorrow}, M.~{Forst},
  S.~{Wall}, X.~{Liu}, M.P.M. {Dean}, arXiv e-prints arXiv:2002.07301 (2020)

\bibitem{dell2016extreme}
M.~Dell'Angela, F.~Hieke, M.~Malvestuto, L.~Sturari, S.~Bajt, I.~Kozhevnikov,
  J.~Ratanapreechachai, A.~Caretta, B.~Casarin, F.~Glerean, et~al., Scientific
  reports \textbf{6}, 38796 (2016)

\bibitem{Mitrano2019ultrafast}
M.~Mitrano, S.~Lee, A.A. Husain, L.~Delacretaz, M.~Zhu, G.~de~la
  Pe{\~n}a~Munoz, S.X.L. Sun, Y.I. Joe, A.H. Reid, S.F. Wandel, G.~Coslovich,
  W.~Schlotter, T.~van Driel, J.~Schneeloch, G.D. Gu, S.~Hartnoll,
  N.~Goldenfeld, P.~Abbamonte, Science Advances \textbf{5}(8) (2019).
\newblock \urlprefix\url{https://advances.sciencemag.org/content/5/8/eaax3346}

\bibitem{Parchenko2019orbital}
S.~Parchenko, E.~Paris, D.~McNally, E.~Abreu, M.~Dantz, E.M. Bothschafter, A.H.
  Reid, W.F. Schlotter, M.F. Lin, S.F. Wandel, et~al., arXiv preprint
  arXiv:1908.02603  (2019)

\bibitem{Kirilyuk2010ultrafast}
A.~Kirilyuk, A.V. Kimel, T.~Rasing, Rev. Mod. Phys. \textbf{82}, 2731 (2010)

\bibitem{Gandolfi2017emergent}
M.~Gandolfi, G.L. Celardo, F.~Borgonovi, G.~Ferrini, A.~Avella, F.~Banfi,
  C.~Giannetti, Physica Scripta \textbf{92}(3), 034004 (2017)

\bibitem{miao2017high}
H.~Miao, J.~Lorenzana, G.~Seibold, Y.~Peng, A.~Amorese, F.~Yakhou-Harris,
  K.~Kummer, N.~Brookes, R.~Konik, V.~Thampy, et~al., Proceedings of the
  National Academy of Sciences \textbf{114}(47), 12430 (2017)

\bibitem{Rohringer2012atomic}
N.~Rohringer, D.~Ryan, R.A. London, M.~Purvis, F.~Albert, J.~Dunn, J.D. Bozek,
  C.~Bostedt, A.~Graf, R.~Hill, et~al., Nature \textbf{481}(7382), 488 (2012)

\bibitem{Beye2013stimulated}
M.~Beye, S.~Schreck, F.~Sorgenfrei, C.~Trabant, N.~Pontius,
  C.~Sch{\"u}{\ss}ler-Langeheine, W.~Wurth, A.~F{\"o}hlisch, Nature
  \textbf{501}(7466), 191 (2013)

\bibitem{Yoneda2015atomic}
H.~Yoneda, Y.~Inubushi, K.~Nagamine, Y.~Michine, H.~Ohashi, H.~Yumoto,
  K.~Yamauchi, H.~Mimura, H.~Kitamura, T.~Katayama, et~al., Nature
  \textbf{524}(7566), 446 (2015)

\bibitem{Wang2017on}
Y.L. Wang, G.~Fabbris, D.~Meyers, N.H. Sung, R.E. Baumbach, E.D. Bauer, P.J.
  Ryan, J.W. Kim, X.~Liu, M.P.M. Dean, G.~Kotliar, X.~Dai, Phys. Rev. B
  \textbf{96}, 085146 (2017)

\bibitem{Basov2017towards}
D.N. Basov, R.D. Averitt, D.~Hsieh, Nature Materials \textbf{16}, 1077 EP
  (2017)

\bibitem{marcus2019cavity}
G.~Marcus, J.~Anton, L.~Assoufid, F.J. Decker, G.~Gassner, K.~Goetze,
  A.~Halavanau, J.~Hastings, Z.~Huang, W.~Jansma, et~al., in \emph{39th Free
  Electron Laser Conf.(FEL'19), Hamburg, Germany, 26-30 August 2019} (JACOW
  Publishing, Geneva, Switzerland, 2019), pp. 282--287

\bibitem{FELOscillator}
K.J. Kim, Y.~Shvyd'ko, S.~Reiche, Phys. Rev. Lett. \textbf{100}, 244802 (2008)

\bibitem{kim2009tunable}
K.J. Kim, Y.V. Shvyd'ko, Physical Review Special Topics-Accelerators and Beams
  \textbf{12}(3), 030703 (2009)

\bibitem{huang2006fully}
Z.~Huang, R.D. Ruth, Physical review letters \textbf{96}(14), 144801 (2006)

\bibitem{maldovan}
M.~Maldovan, Nature \textbf{503}(7475), 209 (2013)

\bibitem{Joe2014emergence}
Y.I. Joe, X.M. Chen, P.~Ghaemi, K.D. Finkelstein, G.A. de~la Pe{\~n}a, Y.~Gan,
  J.C.T. Lee, S.~Yuan, J.~Geck, G.J. MacDougall, T.C. Chiang, S.L. Cooper,
  E.~Fradkin, P.~Abbamonte, Nature Physics \textbf{10}(6), 421 (2014)

\bibitem{Feng2015itinerant}
Y.~Feng, J.~van Wezel, J.~Wang, F.~Flicker, D.M. Silevitch, P.B. Littlewood,
  T.F. Rosenbaum, Nature Physics \textbf{11}(10), 865 (2015)

\end{thebibliography}

\end{document}